\RequirePackage{fix-cm}

\documentclass[smallextended]{svjour3}       

\journalname{Softw Syst Model}

\usepackage{booktabs} 

\usepackage{amssymb} 

\usepackage[acronym]{glossaries} 

\usepackage{subcaption} 

\usepackage[justification=centering]{caption} 

\usepackage{listings}

\usepackage[colorlinks = true,
            linkcolor = blue,
            urlcolor  = blue,
            citecolor = blue,
            anchorcolor = blue]{hyperref}

\usepackage{graphicx}

\def\labelitemi{\textbullet}

\usepackage{xspace}

\newcommand{\eg}{\textrm{e.g.,}\@\xspace}

\newcommand{\ie}{\textrm{i.e.,}\@\xspace}

\newcommand{\etc}{\textrm{etc.}\@\xspace}

\newcommand{\etal}{\textrm{et al.,}\@\xspace}

\newcommand{\tool}{\Simulink Module Tool\@\xspace}%

\newcommand{\Matlab}{\textsc{Matlab}\@\xspace}
\newcommand{\Simulink}{Simulink\@\xspace}

\newcommand\key{\textit}   
\newcommand\ckeyword{\texttt}  
\newcommand\block{\textsf} 
\newcommand\param{\textit}     
\newcommand\file{\texttt}      

\newcommand{\FormatBlocks}{0} 	
\ifthenelse{\equal{\FormatBlocks}{0}}{%
	
\newcommand{\inport}{Inport\@\xspace}
\newcommand{\outport}{Outport\@\xspace}

\newcommand{\subsystem}{Subsystem\@\xspace}
\newcommand{\Subsystem}{Subsystem\@\xspace}

\newcommand{\goto}{Goto\@\xspace}
\newcommand{\from}{From\@\xspace}

\newcommand{\ds}{Data Store\@\xspace}
\newcommand{\dsmem}{Memory\@\xspace}
\newcommand{\dsread}{Read\@\xspace}
\newcommand{\dswrite}{Write\@\xspace}

\newcommand{\bus}{bus\@\xspace}

\newcommand{\model}{Model\@\xspace}

\newcommand{\triggered}{Triggered\@\xspace}

\newcommand{\fcsubsystem}{Function-Call Subsystem\@\xspace}

\newcommand{\simfunc}{Simulink Function\@\xspace}

\newcommand{\simfunccaller}{Function Caller\@\xspace}

\newcommand{\modelref}{Model Reference\@\xspace}
\newcommand{\library}{Library\@\xspace}

\newcommand{\toworkspace}{To Workspace\@\xspace}
\newcommand{\fromworkspace}{From Workspace\@\xspace}

\newcommand{\tofile}{To File\@\xspace}
\newcommand{\fromfile}{From File\@\xspace}

\newcommand{\fromspreadsheet}{From Spreadsheet\@\xspace}

}{

\newcommand{\inport}{\block{Inport}\@\xspace}
\newcommand{\outport}{\block{Outport}\@\xspace}

\newcommand{\subsystem}{\block{Subsystem}\@\xspace}
\newcommand{\Subsystem}{\block{Subsystem}\@\xspace}

\newcommand{\goto}{\block{Goto}\@\xspace}
\newcommand{\from}{\block{From}\@\xspace}

\newcommand{\ds}{\block{Data Store}\@\xspace}
\newcommand{\dsmem}{\block{Memory}\@\xspace}
\newcommand{\dsread}{\block{Read}\@\xspace}
\newcommand{\dswrite}{\block{Write}\@\xspace}

\newcommand{\bus}{\block{bus}\@\xspace}

\newcommand{\model}{\block{Model}\@\xspace}

\newcommand{\triggered}{\block{Triggered}\@\xspace}

\newcommand{\fcsubsystem}{\block{Function-Call Subsystem}\@\xspace}

\newcommand{\simfunc}{\block{Simulink Function}\@\xspace}

\newcommand{\simfunccaller}{\block{Function Caller}\@\xspace}

\newcommand{\modelref}{\block{Model Reference}\@\xspace}
\newcommand{\library}{\block{Library}\@\xspace}

\newcommand{\toworkspace}{\block{To Workspace}\@\xspace}
\newcommand{\fromworkspace}{\block{From Workspace}\@\xspace}

\newcommand{\tofile}{\block{To File}\@\xspace}
\newcommand{\fromfile}{\block{From File}\@\xspace}

\newcommand{\fromspreadsheet}{\block{From Spreadsheet}\@\xspace}

}

\newcommand{\efmodel}{export-function model\@\xspace}

\newacronym{FV}{FV}{Function Visibility}

\newacronym{JMAAB}{JMAAB}{Japan MathWorks Automotive Advisory Board}

\newacronym{MAAB}{MAAB}{MathWorks Automotive Advisory Board}

\newacronym{MISRA}{MISRA}{Motor Industry Software Reliability Association}

\newacronym{MBD}{MBD}{Model-Based Development}

\newacronym{OEM}{OEM}{Original Equipment Manufacturer}

\newacronym{SWCR}{SWCR}{Software Change Request}

\newacronym{MES}{MES}{Model Engineering Solutions}

\newacronym{MCDC}{MCDC}{Modified Condition/Decision Coverage }

\newacronym{SDV}{SDV}{Simulink Design Verifier}

\newacronym{NOP}{NOP}{neutron overpower}

\newacronym{SDS}{SDS}{Shut Down System}

\newacronym{PE}{PE}{Power Estimation}

\usepackage[dont-mess-around]{fnpct}

\newcommand*{\affaddr}[1]{#1}
\newcommand*{\affmark}[1][*]{\textsuperscript{#1}}

\begin{document}

\title{Supporting Modularity in Simulink Models}

\author{Monika Jaskolka \and Vera Pantelic \and Alan Wassyng \and Mark Lawford}

\institute{M. Jaskolka\protect\affmark[1]\affmark[2], V. Pantelic\affmark[1], A. Wassyng\affmark[1], M. Lawford\affmark[1] \at
\affaddr{\affmark[1]
McMaster Centre for Software Certification, McMaster University} \\
\email{\{bialym2, pantelv, wassyng, lawford\}@mcmaster.ca}
\and
\affaddr{\affmark[2]FCA Canada Inc.}\\
\email{monika.jaskolka@fcagroup.com}
}

\date{Received: date / Accepted: date}

\maketitle

\begin{abstract}
\gls{MBD} is widely used for embedded controls development, with \Matlab \Simulink being one of the most used modelling environments in industry. As with all software, \Simulink models are subject to evolution over their lifetime and must be maintained. Modularity is a fundamental software engineering principle facilitating the construction of complex software, and is used in textual languages such as C. However, as \Simulink is a graphical modelling language, it is not currently well understood how modularity can be leveraged in development with \Simulink, nor whether it can be supported with current \Simulink modelling constructs. This paper presents an effective way of achieving modularity in \Simulink by introducing the concept of a \Simulink module. The effectiveness of the approach is measured using well-known indicators of modularity, including coupling and cohesion, cyclomatic complexity, and information hiding ability. A syntactic interface is defined in order to represent all data flow across the module boundary. Four modelling guidelines are also presented to encourage best practice. Also, a custom tool that supports the modelling of \Simulink modules is described. Finally, this work is demonstrated and evaluated on a real-world example from the nuclear domain.
\end{abstract}

\glsresetall 

\section{Introduction}
\label{sec:intro}
\gls{MBD} is a software development approach that uses models to describe the behaviour of a software-intensive application. \gls{MBD} has become an increasingly prevalent paradigm, dominating domains such as aerospace and automotive. \Simulink is one of the most widely used \gls{MBD} platforms, and is the target platform for this research. The benefits of \gls{MBD} include designing at a higher level of abstraction, auto-code generation, simulation, model-in-the-loop testing, \etc In industry, models can become very large and are maintained and evolved over the span of years. This paper explores how to create \Simulink modules that can be used to help make \Simulink designs robust with respect to change. The lack of support for encapsulation and interfaces in \Simulink is also addressed. The paper presents a tool that facilitates building such modules.

We begin by mapping relevant language concepts in \Simulink to C. The C language is widely used in embedded software development, especially by engineers who also develop and maintain \Simulink models---and C does support basic modular encapsulation. We thus draw on C features and coding conventions for modular designs, augmented by knowledge of other programming languages, to give recommendations for achieving modular \Simulink designs.

\emph{Information hiding} is a fundamental principle for modularizing software so that it is \emph{robust with respect to change}~\cite{parnas1972criteria,parnas1985modular}. It aims to decompose a system
such that each likely change (\eg hardware changes, behaviour changes, software design decision changes~\cite{parnas1985modular}) is treated as a ``secret'' and localized (hidden) in a single module. Surprisingly, information hiding and encapsulation have not been readily supported in \Simulink~\cite{bialy2016software,molotnikov2016future}. Parnas criticized the widely used approach of decomposing a system in a ``flowchart'' manner, in which modules are simply major processing steps in the program~\cite{parnas1972criteria}. As \Simulink is a graphical modelling language, this is the de facto method of decomposition currently employed. However, with the introduction of new constructs in the language, in particular \emph{\simfunc{s}}
(available since \Matlab R2014b), it is now possible to design models which break free from the data flow approach. We present a novel approach for decomposing \Simulink models that supports information hiding via the use of \simfunc blocks. In particular, we define a \Simulink module and a module's syntactic interface. Well-defined interfaces are crucial to achieving modularity in designs. A syntactic interface should make clear \emph{all} the communication and dependencies of a module, and ensure that private information is not exposed on the interface. A design pattern to provide a model-level view of this interface is created. We then establish new modelling guidelines to support best practices using the new decomposition and interface concepts, and also provide a tool to support decomposition, interface views, and guideline checking. The proposed approach is  applied to an example problem from the nuclear industry to demonstrate its use and effectiveness. In particular, we seek to objectively evaluate whether the design better supports modularity. We evaluate characteristics that are widely considered to be indicators of design structure and modularity, such as coupling and cohesion, cyclomatic complexity, interface complexity, and testability. The impact to performance is also studied.

This paper is structured as follows. 
Section~\ref{sec:relatedwork} provides an overview of the literature related to model structuring, 
modularity, and interfaces in \Simulink.
Section~\ref{sec:preliminaries} introduces core concepts and explores a relatively new construct in \Simulink, the \simfunc. The relationship between \Simulink constructs and the C language is highlighted.
Section~\ref{sec:module} introduces the idea of a \Simulink module and presents design principles to support modularity. 
Section~\ref{sec:interface} defines the notion of a module interface. A corresponding design pattern is also created in order to represent the interface in a model. 
Tying this all together, Section~\ref{sec:guidelines} presents new guidelines for structuring designs with \simfunc{s} and interfaces. 
Section \ref{sec:tool} introduces an open source tool for supporting module creation, interface representation, and guideline checking. 
Section~\ref{sec:casestudy} demonstrates and evaluates this work on a real-world nuclear example. 
Finally, Section~\ref{sec:conclusions} concludes with a summary and directions for future work.

\section{Related Work}
\label{sec:relatedwork}

In this section, related work on module structure and interfaces is summarized. 

\subsection{Model Structure}

MathWorks is the authority when it comes to \Simulink model structuring. The \Simulink User's Guide~\cite[ch.15]{mathworks2019simulink} provides a guide for choosing an appropriate decomposition construct at the model level, and compares three constructs---\subsystem, \library, and \modelref---according to how each supports the development process, model performance, component reuse, \etc Modularity is not examined explicitly, however, there is a discussion on the related concept of component reuse, which positions the \modelref and \library as well suited for reuse, but not \subsystem{s}. Nevertheless, the ability to hide implementation details is not discussed in the guide. The \Simulink User's Guide also provides recommendations for interface design when it comes to \bus usage, naming conventions, parameter partitioning, and explicit interface configuration~\cite[ch.22]{mathworks2019simulink}.

\begin{figure}
\centering
	\includegraphics[width=.7\columnwidth]{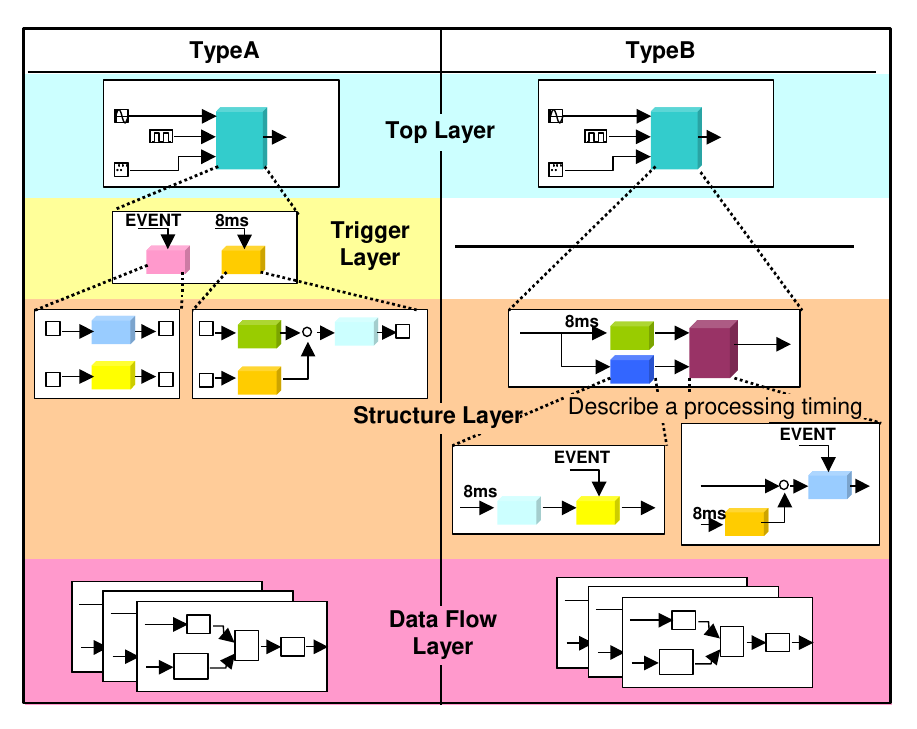}
	\caption{Model structure recommended by \gls{MAAB}~\cite{mathworks2012mathworks}.}
	\label{fig:maab_structure}
\end{figure}

The \gls{MAAB} proposes decomposition using \subsystem blocks also, but more specifically recommends structuring a model into four layers consisting of a root (or top) layer, trigger layer (optional), structure layer, and data flow layer~\cite{mathworks2012mathworks}, as shown in \figurename~\ref{fig:maab_structure}. The root layer gives an overview of the feature being modelled, by showing all model inputs/outputs and their flow into/out of the control \subsystem. The trigger layer is optional, and describes the timing for \triggered and \fcsubsystem{s}. The structure layer is comprised of \subsystem{s}, which organize the control algorithms implemented at the data flow layer. The \gls{MAAB} structure proposes similar ideas that our module structure draws upon. For instance, the top layer should show a complete view of the model inputs/outputs. However, our proposed structure focuses on further decomposition at the structure layer with the intent of enforcing information hiding.

Whalen \etal propose structuring \Simulink models for the purpose of verification~\cite{whalen2014structuring}, decomposing a system into models based on their role: functional, property (requirement), environment, or test input models. Furthermore, vertical decomposition separates each subsystem into its own file. This structure supports independent development and traceability.
		
Dajsuren \etal define metrics for modularity in \Simulink models in terms of coupling (number of exchanged input/output signals) and cohesion (related functionality) for subsystems, ports, and signals~\cite{dajsuren2013simulink}. Metrics include: Coupling Between Subsystems, Degree of Subsystem Coupling, Number of input Ports, Number of output Ports, Number of input Signals, Number of output Signals. Cohesion metrics include: Depth of a Subsystem, Number of Contained Subsystems, and Number of Basic Subsystems. It is clear that the \subsystem is considered to be a module in this context.

In previous work, we performed a thorough comparison of five available \Simulink decomposition constructs for decomposition purposes~\cite{jaskolka2020comparison}. Some of the results are shown in \tablename~\ref{tbl:informationhiding}. The ability to selectively restrict or allow the use of functionality encapsulated by a componentization construct given under ``Limits Use.'' A \simfunc's ability to be scoped makes it unique because one can hide it from other parts of the model in which it resides, or other models. How each construct restricts implicit data flow across its boundaries is also examined. To support information hiding, a construct should ensure that implicitly exposing internal design, or implicitly reading in data, outside of the interface is not possible. We constructed simple experiments to test these scenarios by trying to implicitly pass data across the construct boundaries via \goto/\from blocks and \ds \dsread/\dswrite blocks. A \simfunc prevents hidden data flow by limiting \goto/\from blocks but not \ds \dsread/\dswrite blocks. A full comparison of the various componentization techniques in \Simulink is available~\cite{jaskolka2020comparison}, where the differences between state handling, reusability, and code generation are examined also. As a result, we leverage \simfunc{s} in our proposed module structure in Section~\ref{sec:module}.

\begin{table}
	\centering
	\caption{\Simulink construct support for encapsulation~\cite{jaskolka2020comparison}.}
	\label{tbl:informationhiding}
	\begin{tabular}{lccc}
		\hline
		Construct & Limits Use 	& \multicolumn{2}{c}{Restrict Data Flow}  \\ \cline{3-4}
							&  & Goto/From & Data Store \\ \hline \hline
		Subsystem 									& No	& No 		& No 			\\ \hline
		Atomic Subsystem						& No  & No		& No 			\\ \hline
		Library 										&	No  & No		& No			\\ \hline
		Model Reference 			  		&	No  & Yes		& Local Only \\ \hline
		\Simulink Function 					&	Yes & Yes		& No			\\ \hline
	\end{tabular}
\end{table}

\subsection{Interfaces}

Well-defined interfaces are an integral part of achieving modularity in designs. Commonly, a \Simulink model's interface is considered to be comprised of the \inport{s} and \outport{s} of the top-level system~\cite{doerr2017good,gerlitz2015detection}, also called the \emph{explicit} interface. Bender \etal concluded that \emph{implicit} data flow is a crucial part of a \subsystem's interface, and go on to define a \emph{signature} as a representation of the interface of a \Simulink \Subsystem that effectively captures both the explicit and implicit data flow between \subsystem{s}~\cite{bender2015signature}. We use a similar approach to define a module interface. Rau's work on \Simulink model interfaces recommends simplifying the signal flow into and out of a model using \bus{es} to group them together~\cite{rau2002model}. Masked \subsystem{s} are then used to encapsulate signal operations such as selection, conversion, and renaming. The drawback is a loss of direct visibility of data flow, so we will not incorporate it in our work. The use of pre-/post-condition contracts as verifiable interface specifications for \subsystem{s} has also been proposed~\cite{bostroem2011contract,bostroem2007stepwise,iwu2004practical}. While this design-by-contract approach provides a way of ensuring desired behaviour at the \subsystem-level, our approach aims to document the interface syntax in a complete fashion at the \model-level. Our interface definition can then be used to establish the sets of inputs and outputs between which constraints can be expressed. We discuss interface-related guidelines in Section~\ref{sec:interfaceguidelines}.

In general, the literature sets forth recommendations on the constructs to use for the componentization of models in different contexts and for different purposes, with the most common being \subsystem{s} or \model{s}. No analysis exists that explores which constructs can, and should be, used to support the traditional software engineering principles of modularity and information hiding. 

\section{Core Concepts}
\label{sec:preliminaries}
\begin{figure*}
	\centering
	\begin{subfigure}[b]{.9\columnwidth}
  	\includegraphics[width=\columnwidth]{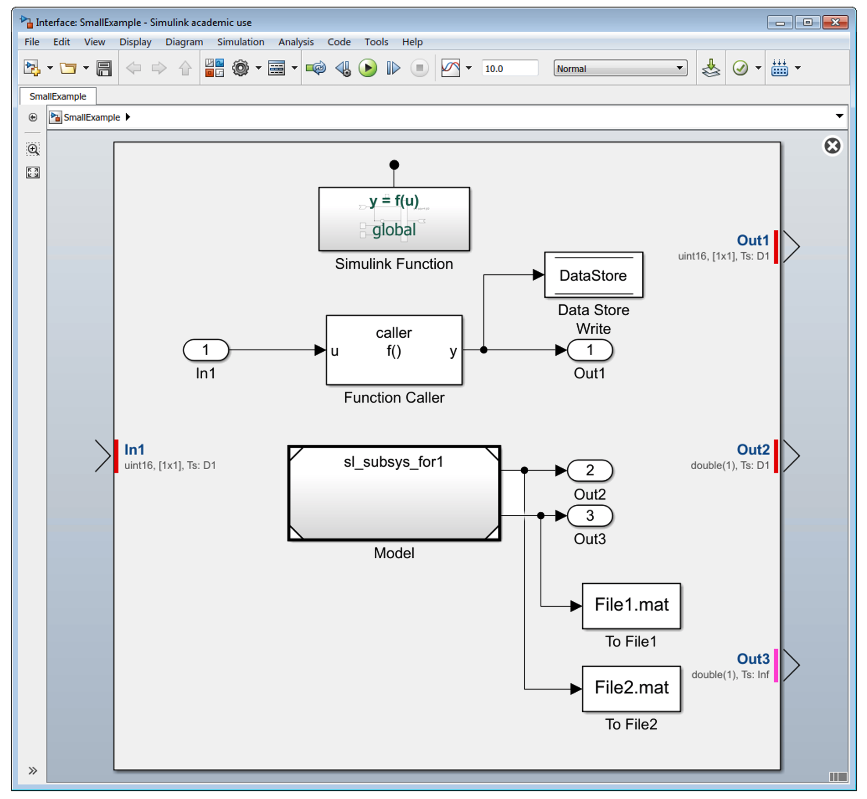}
		\caption{A \Simulink model (in Interface Display view).}
		\label{fig:simulinkinterfacedisplay}
  \end{subfigure}
  \begin{subfigure}[b]{.9\columnwidth}
  	\centering
  	\includegraphics[width=.7\columnwidth]{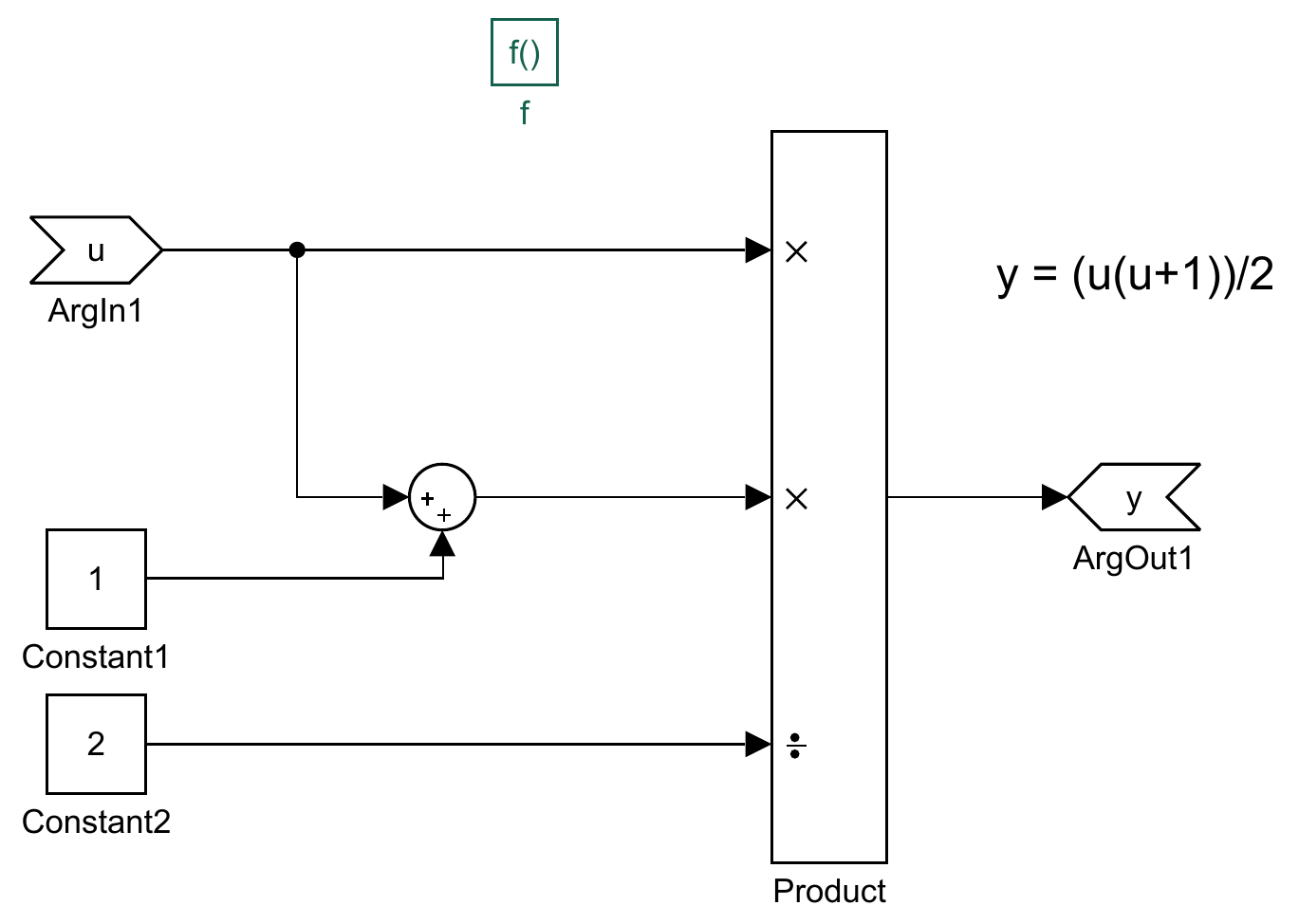}
		\caption{\texttt{\Simulink Function} subsystem from 	\figurename~\ref{fig:simulinkinterfacedisplay}.}
		\label{fig:simpleexample_simfunc}
		\end{subfigure}
	\\[1.25em]
	\begin{subfigure}[b]{.9\columnwidth}
		\centering
  	\includegraphics[width=.8\columnwidth]{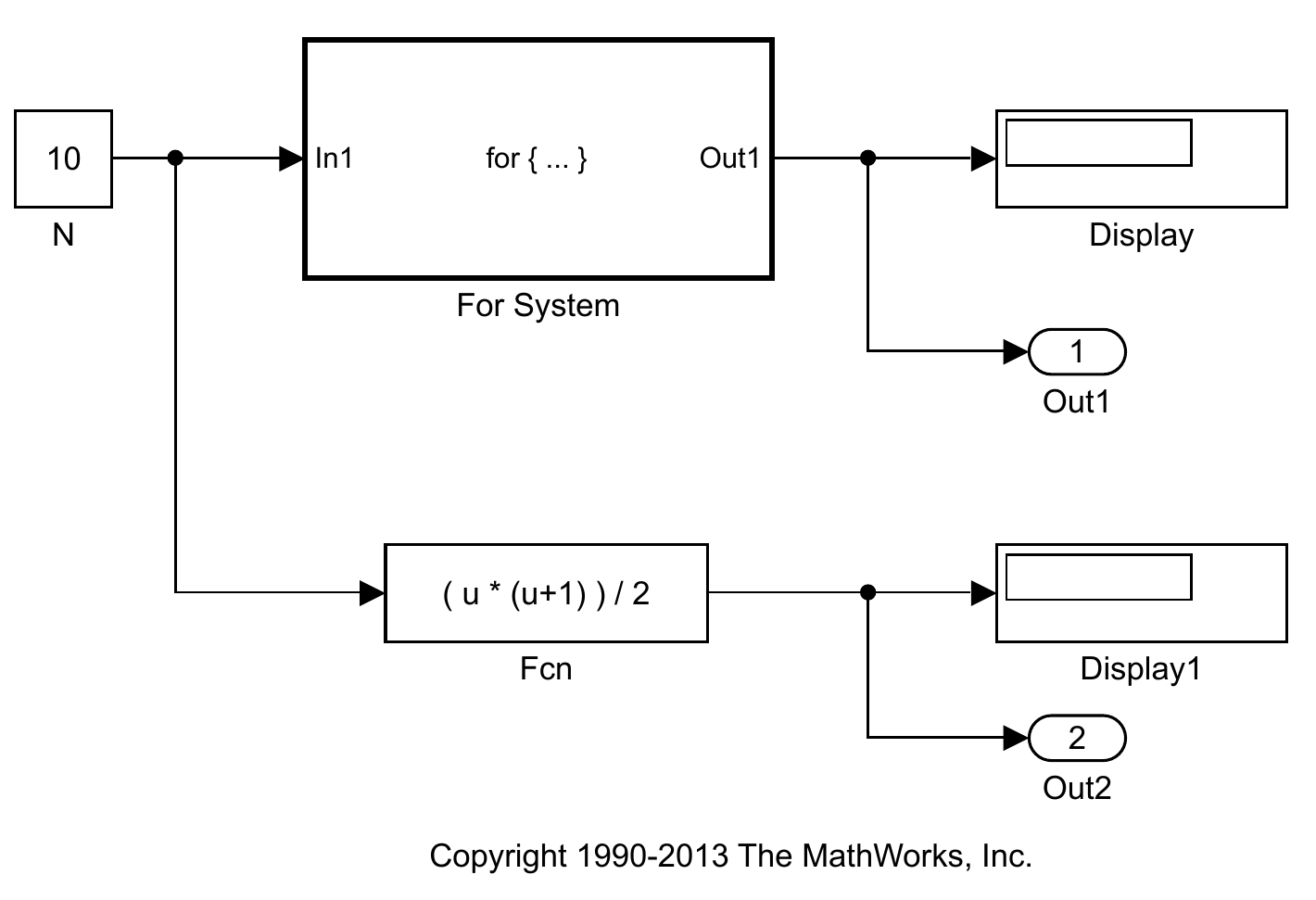}
  	\caption{Model \texttt{sl\_subsys\_for1} referenced in \figurename~\ref{fig:simulinkinterfacedisplay}.}
	\label{fig:simpleexample_modelref}
	\end{subfigure}
  \caption{A simple \Simulink example.}
  \label{fig:simpleexample}
\end{figure*}

In this paper, we investigate modularity principles for \Simulink as a widely-used tool 
for developing embedded control systems. The \Simulink language is comprised of blocks representing different constructs, and each block has parameters that further specify it. 
Blocks are connected via signal lines representing the flow of data or control. Together, these elements form block diagrams known as \emph{models}. \figurename~\ref{fig:simulinkinterfacedisplay} shows an example model that computes the sum of positive integers using three methods. C is the most widely used programming language for embedded software~\cite{aspencore20192019embedded}, and in our experience most developers of \Simulink models also work closely with C. We provide analogies for some \Simulink constructs to the standard C language (C18)~\cite{ieee2018information} to better understand \Simulink and to eventually draw comparisons between their design principles. Both C and \Simulink are not object-oriented languages, enhancing the basis for comparison. Note, we do not discuss how a model is generated into C code, but rather position a model as the primary design artifact in \Simulink, in the same way source files are in C. We are interested in the design-time view, rather than the compile-time view. Thus, a \Simulink model (\file{.mdl}/\file{.slx}) is comparable to a C source file (\file{.c}). However, there is no notion of a header file (\file{.h}) and the interface that it provides. This is due to the top-level block diagram not providing sufficient information about the interface, as explained in Section~\ref{sec:interface}.

\textbf{(Virtual) Subsystems:}
\label{sec:subsystems}
In general, a subsystem is used to group blocks and introduces hierarchical layering. Although there are many kinds of subsystems, the most common is the \emph{virtual} Subsystem.\footnote{We refer to a virtual Subsystem  simply as a \Subsystem.} It assists developers by only visually grouping together blocks. \emph{The \Simulink engine expands virtual \subsystem{s} in place before the execution of the model~\cite{mathworks2019simulink}, akin to a C preprocessor expanding a macro -- but one time only.} We do not elaborate on other nonvirtual (\ie atomic) subsystems besides \simfunc{s}, because they have additional semantics that are not useful for our purposes. 

\textbf{Libraries:}
\label{sec:libraries}
A \library is a special kind of model. Blocks stored in a \library become reusable, as the \library stores the block prototype. Other models can then use an instance of the \library block. Any updates to the \library block will propagate into the models that use it. Instances of a \library block act as references to a library block and are updated/replaced pre-compile time. \emph{Conceptually, a library block is akin to a normal (multi-use) macro in C.}

\textbf{Data Flow:}
\label{sec:dataflow}
Data passing in a \Simulink model is represented using signal lines. However, constructs such as \goto/\from pairs and \ds \dsmem \dsread/\dswrite blocks enable implicit data passing, without a line. This is known as hidden data flow~\cite{bender2015signature} because data can cross certain block boundaries (\eg \subsystem{s}), without being immediately evident. \figurename~\ref{fig:simulinkinterfacedisplay} shows a \ds \dswrite that passes data outside of the model. \emph{In C, variable names represent stored data, while in \Simulink it is mapped out with (named or unnamed) signal lines. In C, there are ways we can store and move data that are difficult to follow; using pointers for instance. Similarly, \Simulink has a variety of ways of storing and moving data that, if not used carefully, make it difficult to understand a model's data flow. For that reason, it is important to define interfaces of modules to improve their understandability.}

\textbf{Workspaces and Data Dictionaries:}
\label{sec:workspaces}
Data and definitions that contribute to the specification of a model can exist outside the block diagram, in the \Matlab (base) workspace, model workspace, and/or one or more data dictionaries~\cite[ch.62]{mathworks2019simulink}. The \Matlab workspace stores temporary data. For permanent data storage, a model workspace is used to save data in a model file, or a data dictionary can store data in a separate \file{.sldd} file.\emph{Workspaces and data dictionaries are similar to a C module dedicated to storing/defining external variables.} 

\textbf{Model References:}
\label{sec:modelref}
A model can be directly referenced from another model using a \modelref block. Unlike libraries, referenced models can be independently simulated. Like subsystems, each \modelref adds hierarchy to the model~\cite[ch.8]{mathworks2019simulink}. \figurename~\ref{fig:simulinkinterfacedisplay} shows a \modelref to the model \texttt{sl\_subsys\_for1}, displayed in \figurename~\ref{fig:simpleexample_modelref}. \emph{The use of a model reference is similar to the C preprocessing directive \ckeyword{\#include}. A referenced model is modelled independently, and is code generated separately from the models that reference it. A model reference in some parent model will make available all exported \Simulink functions (see Section~\ref{sec:simfunc}) to that parent model, in a similar way that including a C header will give access to externally defined functions and variables.}

\textbf{Simulink Functions:}
\label{sec:simfunc}
A \Simulink function is a grouping of reusable logic. It can be defined via a \simfunc block,\footnote{Introduced in R2014b. The \acrlong{FV} parameter was added in R2017b.} or in a Stateflow chart as a \simfunc. As the \simfunc block is the most general of these two, we focus on this kind, and leave the treatment of Stateflow \simfunc{s} to future work. The \emph{scope} of a \simfunc, \ie where it can be used, is determined by both its hierarchical \emph{placement} in the model and its \emph{\gls{FV}} parameter. The \emph{\gls{FV}} parameter can be set to either \emph{scoped} or \emph{global}, but by default it is scoped~\cite{mathworks2019simulink}. The rules for determining the scope of a \simfunc are not straightforward, and are described in detail in~\cite{jaskolka2020comparison}. They are summarized in \tablename~\ref{tbl:summaryFunctionScope} and briefly discussed below.

A \simfunc with global \gls{FV} can be placed anywhere in a model and will be available for external use in the model hierarchy.\footnote{Models exporting functions must be configured as \emph{\efmodel{s}}~\cite[ch.10]{mathworks2019simulink}.} When a \simfunc has scoped visibility, its placement in the model affects its accessibility. If placed at the root it is externally accessible in the model hierarchy. The difference between a global \simfunc and a scoped \simfunc at the root is in the way it is called. In the latter, the function name must be qualified with the \modelref block name. If the scoped \simfunc is placed in a virtual subsystem it is available in the parent subsystem and any descendants; otherwise if placed in a nonvirtual subsystem it is only available at that level.

\begin{table}
	\centering
	\caption{Summary of \simfunc scope.}
	\label{tbl:summaryFunctionScope}
	\begin{tabular}{llll}
		\hline
		Case  & Placement 	& Function\newline Visibility 	& Scope 	\\ \hline \hline
		1 &\emph{Don't care}		& Global 		& External	\\ \hline 
		2 &Root 								& Scoped 		& External	\\ \hline 
		3 &Virtual Subsystem 		& Scoped		& Internal	\\ \hline 
		4 &Nonvirtual Subsystem	& Scoped		& Internal to Subsystem	\\ \hline 
	\end{tabular}
\end{table}

\emph{The concept of a \simfunc is analogous to a function in C, with some semantic differences. While C functions are external by default, \simfunc{s} are scoped by default. In C, one can use functions from a different source by including the header file. To use a \simfunc from a different model, that model must include a \modelref and the function must have external scope. C \ckeyword{static} functions support modularity by restricting the scope of design details. In this case, the function's name is invisible outside of the file in which it is declared, and is analogous to a local \simfunc.}\\

\noindent
A summary of the comparison between C and \Simulink constructs is shown in \tablename~\ref{tbl:summaryComparison}. Given this mapping, \simfunc{s} can be used with other \Simulink constructs to support modularity in a way that facilitates \emph{information hiding}. This is described in the following sections. 

\begin{table}
	\centering
	\caption{Comparison of C and \Simulink constructs.}
	\label{tbl:summaryComparison}
	\begin{tabular}{ll}
		\hline
		C & \Simulink \\ \hline \hline
		Source file & Model \\ \hline
		Header file & \emph{Not Available} \\ \hline
		Include & Model Reference	 \\ \hline
		Function & ``Global" Simulink Function (Case 1) \\ \hline
		Member function & ``Scoped" Simulink Function (Case 2) \\ \hline
		Static function & ``Local" Simulink Function (Case 3 \& 4) \\ \hline
		Macro (single-use) & Virtual Subsystem \\ \hline
		Macro & Library\\ \hline
		Variable & Data Store Memory	\\ \hline
		External data definitions & Workspace/Data Dictionary data \\ \hline
	\end{tabular}
\end{table}

\section{A Simulink Module}
\label{sec:module}
A \emph{module} is a component of a software system. It is a separate unit of a program that encapsulates closely related algorithms (\eg functions, procedures) and data (\eg data structures, variables)~\cite{parnas1972criteria}. Encapsulation means restricting access to a portion of the module, such that certain elements are not accessible outside of the module, but can be manipulated via public elements revealed on the module's interface. In this section, the notion of a module in \Simulink is introduced, drawing from the C analogy in Section~\ref{sec:preliminaries}.

\begin{figure}
\centering
	\includegraphics[width=.8\columnwidth]{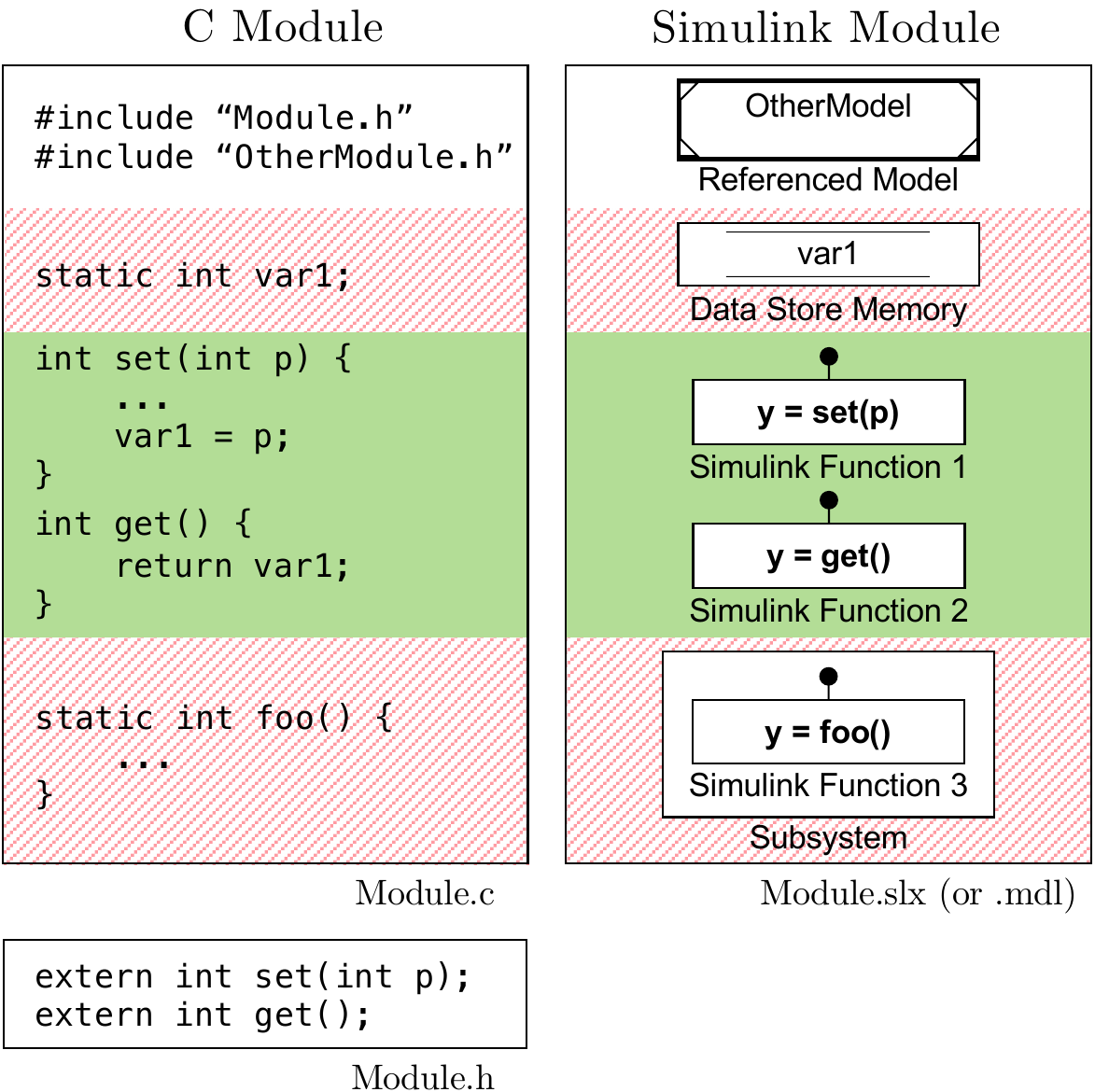}
	\caption{Module structure in \Simulink based on C.}
	\label{fig:module}
\end{figure}

Modular programming in C entails decomposing a system into separate modules~\cite{srivastava2008modular,oualline1997practical}. Each module consists of a source file that groups together definitions of related functionality and data, while a module's interface is described by its header file. A module's implementation should be considered private, or internal, to the module, and only those elements listed on the interface are accessible to other modules which import the interface. The ability to selectively hide or expose functions is achieved via the use of the \ckeyword{static} keyword, as shown in \figurename~\ref{fig:module}, where the functions \ckeyword{set} and \ckeyword{get} are public, while the static function \ckeyword{foo} and static variable \ckeyword{var1} are private. If another module wishes to use this module's public elements, it can do so by including the module's interface, and then making calls to public functions in the module, with parameter values from the calling program.

It is possible to employ the same modular approach in \Simulink. Although there are several componentization techniques in \Simulink, we wish to utilize the construct that best supports encapsulation, and thus information hiding. 
In previous work we found that a \simfunc's unique ability to be scoped enables one to effectively hide it from other parts of the model in which it resides, or other models (\tablename~\ref{tbl:informationhiding}). Moreover, a \simfunc helps to prevent some hidden data flow implicitly crossing over its boundary. \simfunc{s} prove to be best suited to help us build modular \Simulink designs that can actively support encapsulation and thus facilitate information hiding. The result of the comparison of \Simulink componentization techniques (\ie \cite{jaskolka2020comparison}) leads to the proposed method for constructing modules in \Simulink. \figurename~\ref{fig:module} illustrates how we can build \Simulink modules based on the essential components of a module in C. The \Simulink module uses: i) \emph{Model References} to import other \Simulink modules; ii) properly scoped \emph{Data Stores} as state data private to the module; iii) \emph{Simulink Functions} as functions exported by the module; and iv) \emph{Subsystems} to restrict a \simfunc so that it is private to the module. Like in C, the proposed \Simulink modules are not object-oriented classes, and cannot be instantiated multiple times to create multiple objects. However, they make it possible to achieve \emph{information hiding} in \Simulink designs, so that the designs become more robust with respect to (foreseen) changes.

\section{\Simulink Module Interfaces}
\label{sec:interface}
An interface is the set of services that each module provides to its clients~\cite{ghezzi2002fundamentals}. A syntactic interface is generally represented as a statement of elements and their properties that the module chooses to make known to a user or client modules. As shown in \figurename~\ref{fig:module}, a \Simulink module has no concept of an explicit interface like that provided in C. For this reason, we now define a \Simulink module interface, so as to be able to extract it automatically from a \Simulink module. A syntactic module interface contains:

\begin{itemize}
\renewcommand{\labelitemi}{$\bullet$}
	\item inputs --- data the client needs to provide 
	\item outputs --- data the module promises to provide
	\item exports --- functionality the module provides to users
\end{itemize} 

It is important that no unnecessary information is disclosed to the client on the interface. The client needs to know what the module agrees to provide via the interface, but does not need to understand the details of the implementation. As long as the interface remains the same, changes to the implementation can take place without affecting users in any way. 

This interface is consistent with how many programming languages specify interfaces. Modern programming languages typically use keywords such as \emph{Definition} or \emph{Public} to delineate the ``interface'', and \emph{Implementation} or \emph{Private} to separate out the private implementation. The interface usually consists of only those elements that are necessary to make use of the exported functionality (\eg constant, type, variable, function prototype), but can also import elements that are needed in the interface itself. Note that it is usually possible to make module variables public, but in the spirit of information hiding, module variables (as opposed to ``parameters'') should never be exposed on the interface.

The prevailing view is that a \Simulink \emph{model's} interface is comprised of the \inport{s} and \outport{s} of the top-level system (\eg \cite{doerr2017good,gerlitz2015detection}). This is reflected in the \emph{Interface Display} feature provided by \Simulink, as shown in \figurename~\ref{fig:simulinkinterfacedisplay}. The Interface Display aims to provide users with a better view of the system's interface~\cite[ch.12]{mathworks2019simulink}, and we can see that one \inport and three \outport{s} are displayed around the edge of the model. However, a model's interaction with other systems and its environment can consist of additional communication constructs that Interface Display fails to show---in this case a \ds \dswrite, two \tofile blocks, and an exported \simfunc. Interface Display is thus insufficient in describing the actual interface of a model. Although this example is simple, the interface of a model can quickly become more difficult to understand due to the fact that these constructs can be placed \emph{anywhere} in a model, potentially several layers deep. Thus, it can be difficult to ``see'' a \Simulink module's interface and to understand what the module is exposing to other modules. The notion of \emph{signature} of a \Simulink \subsystem can be used to represent the subsystem's interface~\cite{bender2015signature}. It addresses the concerns with implicit data flow between subsystems by including in the subsystem's signature both the explicit data flow mechanisms (\ie \inport/\outport) and the implicit data flow mechanisms (scoped or global \goto/\from blocks and \ds \dsread/\dswrite/\dsmem blocks). We build on this idea to define the module's interface and represent it in the module in a similar way. This definition was then used to develop a design pattern in \Simulink, as well as tool support to automatically extract and visualize it in a model.

\subsection{Definition}
\label{sec:interfacedef}
\begin{figure}
\centering
		\includegraphics[width=.8\columnwidth]{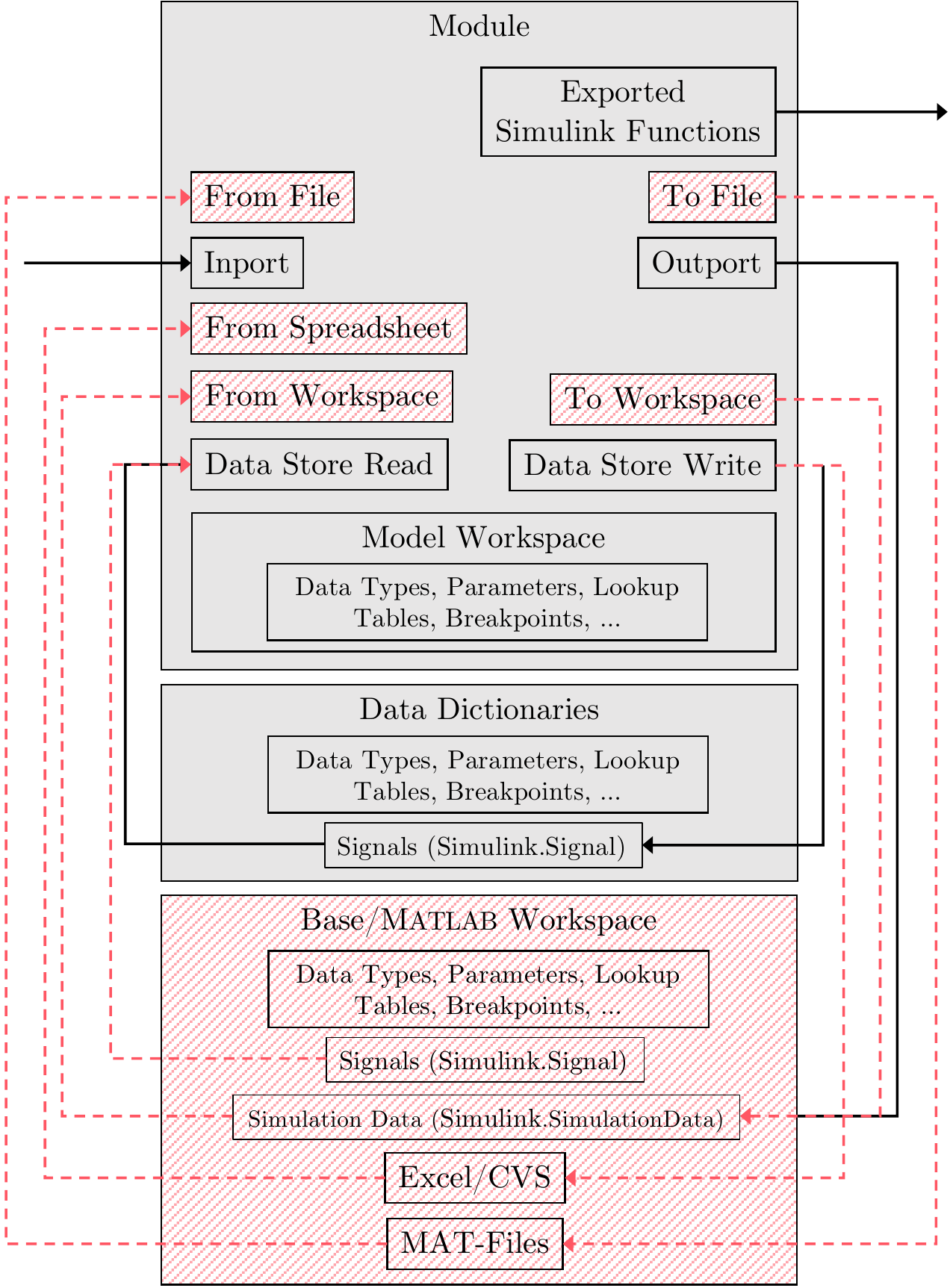}
		\caption{All possible interface data flow,\\with restricted items dashed/crossed out.} 
	\label{fig:simulinkinterfaceflow_all}
\end{figure}

A depiction of all the elements of a \Simulink module's interface is shown in \figurename~\ref{fig:simulinkinterfaceflow_all}. Shared data is defined via \inport blocks, as well as others that are usually not considered, such as the \fromfile and global \ds{s} that are read. The interface also includes the module outputs via \outport blocks, as well as other blocks such as \toworkspace and global \ds{s} that are written to. Shared functions are \simfunc{s} that are exported from a module. The lines (both solid and dashed) show all the possible data flow between a module and other workspaces. It is our recommendation to restrict the flow of data, as denoted by the dashed/crossed out items, however, this is discussed in Guideline 4 in Section~\ref{sec:guidelines}.

\newcommand{\nc}{\newcommand}

\nc{\Parent}{\operatorname{\textsl{parent}}}
\nc{\Children}{\operatorname{\textsl{children}}}
\nc{\AtRoot}{\operatorname{\textsl{atRoot}}}
\nc{\BlockSet}{\mathcal{B}(M)}
\nc{\SubsystemSet}{\mathcal{S}(M)}

\nc{\Inport}{\operatorname{\textsl{IP(M)}}}
\nc{\RootInport}{\operatorname{\textsl{IR(M)}}}
\nc{\rootinport}{\operatorname{\textsl{ir}}}

\nc{\Outport}{\operatorname{\textsl{OP(M)}}}
\nc{\RootOutport}{\operatorname{\textsl{OR(M)}}}
\nc{\rootoutport}{\operatorname{\textsl{or}}}

\nc{\ModelRef}{\operatorname{\textsl{MR(M)}}}
\nc{\LinkedLib}{\operatorname{\textsl{LL(M)}}}

\nc{\FcnDef}{\operatorname{\textsl{FD(M)}}}
\nc{\FG}{\operatorname{\textsl{FG(M)}}}

\nc{\FS}{\operatorname{\textsl{FS(M)}}}

\nc{\FL}{\operatorname{\textsl{FL(M)}}}
\nc{\fl}{\operatorname{\textsl{fl}}}

\nc{\DataStore}{\textnormal{\textsl{DS(M)}}}
\nc{\DataStoreWrite}{\textnormal{\textsl{DSW(M)}}}
\nc{\DataStoreRead}{\textnormal{\textsl{DSR(M)}}}

\nc{\ToWorkspace}{\textnormal{\textsl{TW(M)}}}
\nc{\FromWorkspace}{\textnormal{\textsl{FW(M)}}}
\nc{\ToFile}{\textnormal{\textsl{TF(M)}}}
\nc{\FromFile}{\textnormal{\textsl{FF(M)}}}
\nc{\FromSpreadsheet}{\textnormal{\textsl{FS(M)}}}


We can describe a \Simulink module interface via standard set-theoretic definitions. These definitions not only help to make the interface precise, they also support the creation of tools. In particular, we use this definition to implement the interface extraction and representation in the \tool. We define a \Simulink module $M$ simply as a set of blocks. We consider the block diagram (top-level) system itself to be a part of this set, so as to treat the system and subsystems the same. We abstract away the notion of signal flow between the blocks, as we are not concerned with intra-module communication (it was previously addressed~\cite{bender2015signature}), but rather the inter-module communication not represented by conventional signals.

\begin{definition}[Identifiers I]
\begin{itemize}
	\item[]
	\item $\BlockSet$ is the set of \emph{all} blocks in the module $M$, \ie those at the top-level as well as any that are contained within other blocks (regardless of hierarchy).
	\item $\SubsystemSet$ is the set of all \Subsystem blocks in the module $M$, as well as the root system, so $\SubsystemSet \subseteq \BlockSet$.
\end{itemize}
\end{definition}

\begin{definition}[Block Containment]
	For some blocks $b$ and $c$, $b \in c$ denotes that $b$ is wholly contained in $c$. It can also be said that $b$ is a child of the container $c$.
\end{definition}
  
\begin{definition}[Parent Block]
	The partial function $\Parent: \BlockSet \rightarrow \SubsystemSet$ is defined,
\[
\Parent(b) =
  \begin{cases} 
  	s 									& s \in \SubsystemSet \land b \in s \\
  	\mathsf{undefined} 	& \text{otherwise}
  \end{cases}
\]
\end{definition}

\begin{definition}[Root Block]
	The function $\AtRoot: \BlockSet \rightarrow \mathbb{B}$, where $\mathbb{B} =\{\mathsf{false},\mathsf{true}\}$, \ is defined,
\[
\AtRoot(b) =
  \begin{cases} 
  	\mathsf{true}		& \Parent(\Parent(b)) = \mathsf{undefined} \\
  	\mathsf{false} 	& \Parent(\Parent(b)) \in \SubsystemSet
  \end{cases}
\]
A block $b$ is in the root system of module $M$ when its parent in turn does not have a parent, or, block $b$ has no defined grandparent.
\end{definition}

\begin{definition}[Identifiers II]
\begin{itemize}
		\item[]
		\item $\Inport$ is the set of all \inport blocks
		\begin{itemize}
				\item $\RootInport$ represents root-level inports ($ \RootInport \subseteq \Inport$) 		and is defined,	$\RootInport = \{\rootinport \mid \rootinport \in \Inport \land \AtRoot(\rootinport)\}$
		\end{itemize}
		
		\item $\Outport$ is the set of all \outport blocks
		\begin{itemize}
				\item $\RootOutport$ represents root-level outports ($ \RootOutport \subseteq \Outport$)  and is defined, $\RootOutport = \{\rootoutport \mid \rootoutport \in \Outport \land \AtRoot(\rootoutport)\}$
		\end{itemize}

  	\item $\FcnDef$ is the set of all \simfunc blocks
		\begin{itemize}
			\item $\FG$ represents global functions ($\FG \subseteq \FcnDef$) 
			\item $\FS$ represents scoped functions ($\FS \subseteq \FcnDef$) 
			\item $\FL$ represents local functions ($\FL \subseteq \FS$) and is defined, $\FL = \{\fl \mid \fl \in \FS \land \neg\AtRoot(\fl)\}$
		\end{itemize}
		
		\item $\ToFile$ is the set of all \tofile blocks
		\item $\FromFile$ is the set of all \fromfile blocks
		\item $\FromSpreadsheet$ is the set of all \fromspreadsheet blocks
		
		\item $\ToWorkspace$ is the set of all \toworkspace blocks
		\item $\FromWorkspace$ is the set of all \fromworkspace blocks
 
  	\item $\DataStore$ is the set of all global data stores 
		\begin{itemize}
			\item $\DataStoreRead$ represents global data stores that have a corresponding \ds \dsread  block
			\item $\DataStoreWrite$ represents global data stores that have a corresponding \ds \dswrite  block
		\end{itemize}
\end{itemize}
\end{definition}

\begin{definition}[Inputs]
\label{def:inputs}
$IN(M)$, the inputs of module $M$, is a tuple ($I$, $F$, $S$, $W$, $D$) of root-level \inport, \fromfile, \fromspreadsheet, \fromworkspace, and global \ds \dsread blocks, where $I = \RootInport$, $F = \FromFile$, $S = \FromSpreadsheet$, $W = \FromWorkspace$ and $D = \DataStoreRead$. 
\end{definition}

\begin{definition}[Outputs]
\label{def:outputs}
$OUT(M)$, the outputs of module $M$, is a tuple ($O$, $F$, $W$, $D$) of root-level \outport, \tofile, \toworkspace, and global \ds \dswrite blocks, where 
 $O = \RootOutport$, $F = \ToFile$, $W = \ToWorkspace$, and $D = \DataStoreWrite$. 
\end{definition}

\begin{definition}[Exports]
\label{def:exports}
$EX(M)$, the exports of module $M$, is the set $E$ of exported \simfunc blocks, 
    $E = \FG \cup (\FS \setminus \FL)$.
Global \simfunc{s} are included as they are always on the module interface. Scoped \simfunc{s} are included if they are at root-level, \ie they will be exported on the module interface. 
\end{definition}

\begin{definition}[Interface]
\label{def:interface}
	$\mathcal{I}(M)$, the interface $\mathcal{I}$ of a module $M$, is a tuple of inputs, outputs, and exports, 
		$\mathcal{I}(M) = (IN(M), OUT(M), EX(M))$.
\end{definition}

\subsection{Design Pattern}
\label{sec:designpattern}
\begin{figure}[htb]
\captionsetup{justification=centering}
\begin{subfigure}[b]{.43\columnwidth}
	\includegraphics[width=.9\columnwidth]{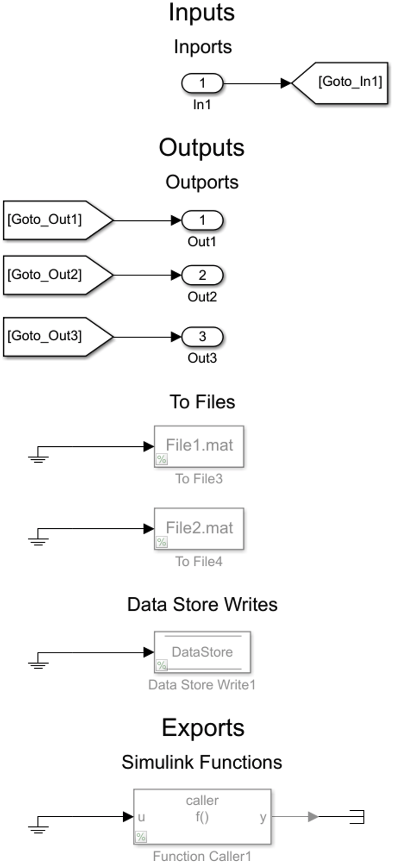}
	\caption{Visual interface representation via design pattern.}
	\label{fig:simulinkinterface_visual_dev}
\end{subfigure}
~
\begin{subfigure}[b]{.48\textwidth}
	\lstset{basicstyle=\footnotesize\ttfamily, keepspaces=false, columns=flexible,}
	\begin{lstlisting}
Inputs
------
Inports:
  SmallExample/In1, uint16, 1, 1

Outputs
-------
Outports:
  SmallExample/Out1, Inherit: auto, 1, 1
  SmallExample/Out2, Inherit: auto, -1, 1
  SmallExample/Out3, Inherit: auto, -1, 1
To Files:
  SmallExample/To File1, Timeseries, N/A, -1
  SmallExample/To File2, Timeseries, N/A, -1
Data Store Writes:
  SmallExample/Data Store Write, uint16, 1, 1

Exports
-------
Simulink Functions:
  SmallExample/Simulink Function, 
    In:  uint16, 1, -1
    Out: uint16, 1, -1
\end{lstlisting}
\caption{Textual interface description.}
\label{fig:simulinkinterface_text}
\end{subfigure}
\caption{Interface representations for \figurename~\ref{fig:simulinkinterfacedisplay},\\as generated by the \tool.}
\end{figure}

We can now create a design pattern within the model file to provide an easily understood view of the module's interface based on Definition~\ref{def:interface}. The design pattern is placed in a \Simulink module at the root, to the left of any other elements at that hierarchical level. It contains labelled sections corresponding to Definitions~\ref{def:inputs}--\ref{def:exports}. Where possible, the interface design pattern is represented using commented out blocks, thus preventing it from having any behavioural impact on the module or adding new code during code generation. If instantiating the design pattern in the module is not possible, the interface can be represented in text form in the \Matlab Command Window. In the textual description of the interface, each element's full path name, data type, dimensions, and sample time are listed. The tool we developed, the \tool (Section~\ref{sec:tool}), supports the automatic creation of an interface in both visual and textual forms. The visual interface for \figurename~\ref{fig:simulinkinterfacedisplay}, as generated by the tool, is shown in \figurename~\ref{fig:simulinkinterface_visual_dev}, and has four elements that the MathWorks Interface Display does not show. This is a concrete application of the definition of a module interface as presented in Section~\ref{sec:interfacedef}. 

\subsection{Benefits}
\label{sec:interfacebenefits}
There are several practical benefits and situations in a software engineering methodology where an interface in a \Simulink module is beneficial. We describe these use cases in what follows.

\paragraph{Development} Passing information that is too detailed, unnecessary, arbitrary, or potentially changeable, violates software design principles. Clear interfaces help developers review them critically and examine whether their constituents are, for example, likely to change, too low-level, or unnecessary.

\paragraph{Collaboration} The presence of an interface is also invaluable in understanding a module for the first time, particularly when it originates from a different developer or source. If an interface is provided, the developer can use the module in a black-box fashion, without taking the time to understand the internals of the design. 

\paragraph{Testing} 
With an easy to identify interface, all the module inputs and outputs are evident to a tester. When using a third-party testing tool such as Reactis by Reactive Systems, a developer can quickly identify module inputs/outputs that may not be included automatically by the tool. In such cases, a complete and visual interface flags them and allows the user to deal with them appropriately, resulting in better coverage~\cite{bender2015signature}.

\paragraph{Production}
Several constructs that can be on a module interface are not recommended for a module that is to be used to generate production code (\eg \tofile, \fromworkspace, \etc). However, these constructs are useful to developers during module development and simulation. An interface will capture such constructs, and empower developers to use them with the knowledge that they will be easy to identify and remove once a module is ready to be transitioned to production. 

\paragraph{Documentation, Refactoring, and Maintenance}
Documentation of \Simulink models is often deficient~\cite{schaap2018documenting,pantelic2019something}. It can be difficult for developers to understand the overall functionality of complex models, as well as how they interact with other models. An interface makes this clear at the root level of the module, saving the developer from navigating to other levels. Structuring a module such that it always contains an up-to-date interface eases documentation efforts, and supports the concept of ``self-documenting'' software. 

\section{Modelling Guidelines}
\label{sec:guidelines}
The use of \simfunc{s} and a syntactic interface help create \Simulink modules. Guidelines are useful for further supporting good practices when using these approaches. In this section, we discuss existing modelling guidelines for \Simulink, and present new ones to address gaps where current guidelines fall short. The \tool provides automated compliance checking for these guidelines, and is discussed in Section~\ref{sec:tool}. A user is able to select one or more of these guidelines, and any violating blocks will be reported. To the best of our knowledge, no other tools support these guidelines.

\subsection{Simulink Functions}
MathWorks is the de facto authority on best practices for designing with \Simulink. Their advisory boards~\cite{mathworks2012mathworks,mathworks2015japan} provide the most influential guidelines, but currently, none address \simfunc scoping. The \gls{MISRA} \Simulink guidelines also pre-date \simfunc{s}~\cite{misra2009misra}. Recommendations on using \simfunc{s} in order to promote best practices for supporting modularity and information hiding are introduced below. These guidelines are widely accepted in other languages to increase understandability, promote maintainability, and reduce errors, thus, we adapt them for \Simulink R2014b and newer releases.

%

\vspace{.5em}
\noindent\textbf{Guideline 1 (\Simulink Function Placement)} \textit{
Place the \simfunc block in the lowest common parent of its corresponding \simfunccaller blocks. Do not position the \simfunc in the top layer for no reason. Avoid placing \simfunc blocks below their corresponding \simfunccaller blocks.}

\vspace{.5em}
\noindent\textbf{Guideline 2 (\Simulink \acrlong{FV})} \textit{
Limit the \param{\gls{FV}} parameter of the \simfunc block's trigger port to \param{scoped} if possible.}
\vspace{.5em}

In textual programming languages, it is good practice to ensure variables and functions are declared at the minimum scope from which their identifiers can still reference them~\cite{martin2008clean}. This promotes readability, reliability, and reusability of the code~\cite{unicarnegiemellon2020sei}. In \Simulink, the same treatment is recommended for \ds \dsmem blocks and \goto blocks, in order to support code comprehension, maintenance, as well as to avoid unintended access~\cite{pantelic2018software,mathworks2012mathworks,misra2009misra}. For these reasons, we introduce the two aforementioned guidelines for \simfunc blocks.

Guideline 1 describes how to position \simfunc blocks in order to minimize their accessibility both inside and outside the module. This is achieved by placing \simfunc blocks as low as possible in the hierarchy, while still allowing any function calls to reference their corresponding \simfunc block without added name qualifiers. The exception to this occurs when the intent is to associate a \simfunc with its parent subsystem. This may be to increase the reusability of the subsystem itself, so the \simfunc is encapsulated by that subsystem, even though \simfunccaller{s} may be present above it in the hierarchy.

The hierarchical placement of a \simfunc can also affect its presence on the module's interface. If it is \param{scoped} and placed at the root, it will be externally accessible by other modules. A similar treatment for a \simfunc's \param{\gls{FV}} parameter is recommended in Guideline 2. It should be set to its most restrictive setting if possible.


\vspace{.5em}
\noindent\textbf{Guideline 3 (\Simulink Function Shadowing)} \textit{
Do not place \simfunc{s} with the same name and input/output arguments within each other's scope.}
\vspace{.5em}

Function overloading occurs when multiple definitions of a function exist with the same name but different input or output arguments. \Simulink does not allow a \simfunc to be placed in a \subsystem that already contains a \simfunc with the same name. However, if the placement of a \simfunc is at a different hierarchical level than another of the same name, one can define functions that shadow/mask each other. However, since scoping rules for \simfunc{s} are complex, and users may be unaware of a naming collision, it is best to avoid situations where more than one function with the same name and arguments is accessible. The \gls{JMAAB} guideline jc\_0791 recommends a similar treatment for data stored in multiple workspaces~\cite{mathworks2019simulinkb}.

\subsection{Interfaces}
\label{sec:interfaceguidelines}
The \Simulink User's Guide discusses good practices for interface design, including \Simulink subsystem interfaces~\cite{mathworks2019simulink}. The guidelines provide information about where model objects/data \emph{can} be stored, but provide no real guidance on where they \emph{should} be stored. Moreover, the use of constructs that contribute to hidden data flow into or out of the model is not addressed. The MathWorks \Simulink Check provides guidelines for ``high integrity systems modelling'' for models that must comply with DO-178C/DO-331, ISO 26262, and other standards~\cite{mathworks2019simulinkd}. 
One guideline recommends that top-level \inport blocks must have \param{data type}, \param{port dimensions}, and \param{sample time} parameters populated. This is good practice in general and will assist in making the details of the interface data flow clear. 

\gls{MAAB}/\gls{JMAAB} provide a single guideline regarding interfaces, recommending the enabling of strong-typing in Stateflow charts. This is not directly useful for examining the model's top-level interface. \gls{MAAB}/\gls{JMAAB} also provides a guideline which lists prohibited \Simulink blocks, including \tofile and \toworkspace blocks in control models~\cite{mathworks2012mathworks,mathworks2015japan}. Similarly, the Embedded Coder User's Guide describes which \Simulink blocks support C code generation, and provides details on how certain blocks are treated during code generation~\cite[ch.2]{mathworks2019embedded}. In particular, the blocks described in our proposed definition of an interface are treated as follows.

\begin{itemize}
	\item \emph{Supported}: \inport/\outport, \ds \dsread/\dswrite, \modelref, \library, \simfunc, \simfunccaller
	\item \emph{Ignored}: \toworkspace/\fromworkspace
	\item \emph{Not recommended for production}: \tofile/\fromfile, \fromspreadsheet
\end{itemize}

Although \tofile, \fromfile, and \fromspreadsheet blocks are not recommended for \emph{production}, developers may use them during \emph{development} because they are valuable for prototyping and logging purposes. Thus, \tofile, \fromfile, and \fromspreadsheet should be represented on the interface. When using these blocks for prototyping, an interface that highlights these constructs will help in identifying them so they can be removed once the design is finalized. This approach will help support the Embedded Coder guideline.


\vspace{.5em}
\noindent\textbf{Guideline 4: (Use of the Base Workspace)} \textit{
Do not use the base workspace for storing, reading, or writing data that a module is dependant on. Instead, place data in either the model workspace, if it is used in a single module, or a data dictionary if it is shared across modules.}
\vspace{.5em}

A likely change for a module that is used for code generation is that it will change workspaces, from being situated in the base workspace of the \Simulink development environment, to being flashed onto the target embedded device. One can anticipate and prepare for this future change by creating a stable interface from the first stages of development. This not only minimizes the need for changes later on, but can also reduce dependencies. This is achieved by restricting the use of interface elements that are used for prototyping or do not support code generation. In particular, developers should avoid using the base workspace for storing, reading, or writing data (including types, signals, \etc). 

To see Guideline 4 applied, \figurename~\ref{fig:simulinkinterfaceflow_all} has restricted items dashed/crossed out, such as the base workspace and associated constructs. As a result, the data flow has been simplified significantly, with the dashed lines showing data flow that is eliminated. Interestingly, \gls{MAAB} explicitly prohibits the use of \tofile and \toworkspace blocks, but recommendations for their counterparts, the \fromfile and \fromworkspace blocks as well as the \fromspreadsheet block, are not provided.

\section{The \tool}
\label{sec:tool}

\begin{figure}
\centering
	\begin{subfigure}[b]{\columnwidth}
		\centering
		\includegraphics[width=\columnwidth, trim={3em 2em 0em 6em}, clip]{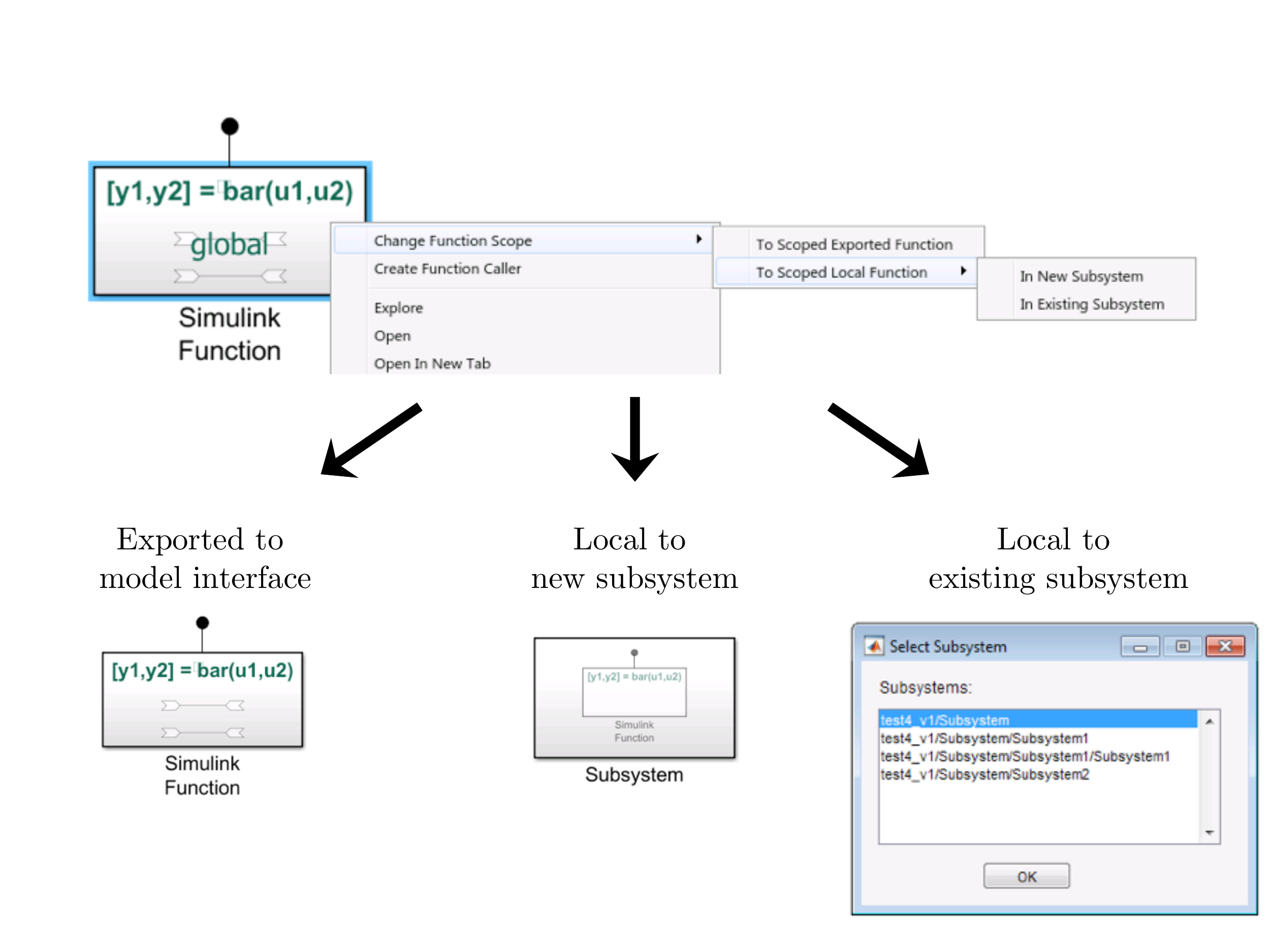}
		\caption{Changing the scope of a \simfunc.}
		\label{subfig:scope}
	\end{subfigure}
	\begin{subfigure}[b]{\columnwidth}
		\centering
			\includegraphics[width=\columnwidth, trim={4em 18em 3em 19em}, clip]{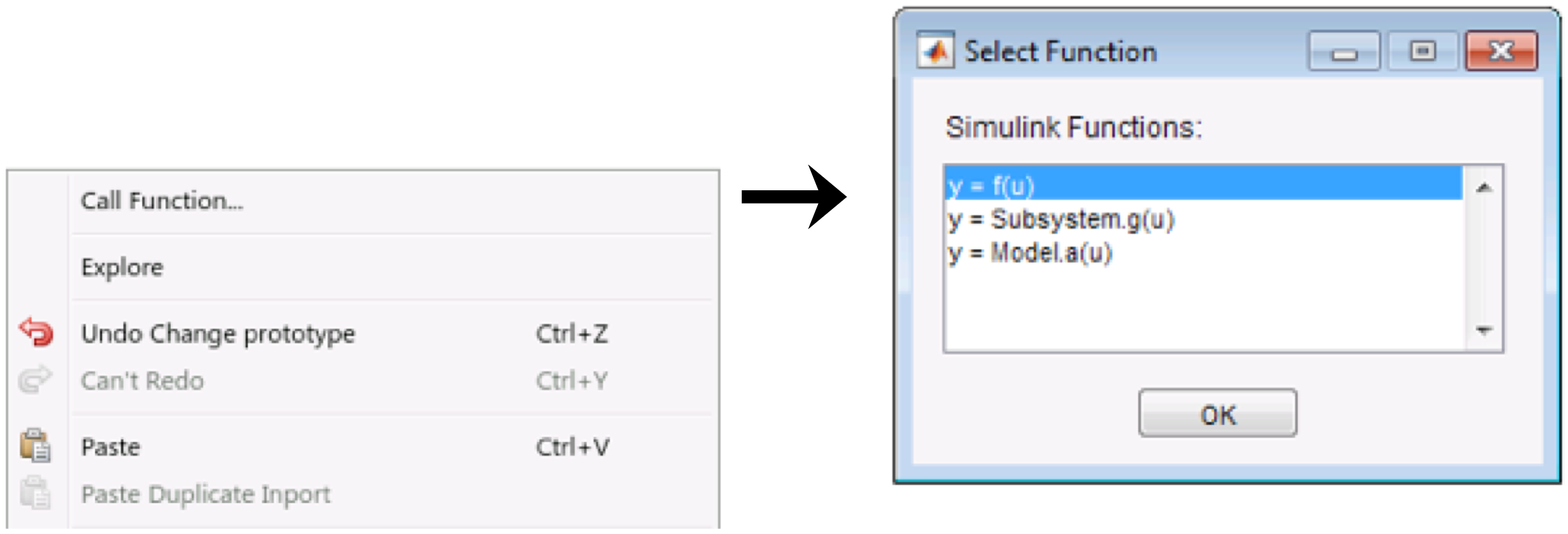}
		\caption{Calling \simfunc{s} that are in scope.}
		\label{subfig:call}
	\end{subfigure}
	\vspace{1em}
	
	\begin{subfigure}[b]{\columnwidth}
		\centering
			\includegraphics[width=.45\columnwidth]{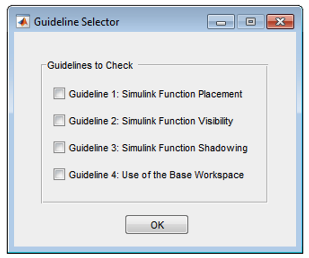}
		\caption{Checking module guideline compliance.}
		\label{subfig:guideline}
	\end{subfigure}
	\caption{\tool.}
	\label{fig:simulinkhelper}
\end{figure}

We developed the \tool to assist with applying our approach for constructing \Simulink modules as described in Section~\ref{sec:module}, generating the interface defined in Section~\ref{sec:interface}, and checking compliance to the guidelines proposed in Section~\ref{sec:guidelines}. It is open-source and available on the \Matlab Central File Exchange.\footnote{\url{www.mathworks.com/matlabcentral/fileexchange/71952-simulink-module-tool}} The capabilities of the tool are as follows:

\begin{itemize}
\item The tool converts between the different kinds of scoping for \simfunc{s} (\figurename~\ref{subfig:scope}), so the user does not have to be concerned with remembering the complex scoping rules regarding \param{\gls{FV}} and placement. 
\item The tool assists users in calling \simfunc{s} that are in scope, with their appropriate qualifiers (\figurename~\ref{subfig:call}). Right-clicking in the model and then selecting \key{Call Function...} from the Context Menu displays a listbox showing \simfunc{s} that can be called from that location. Making a selection from this list creates a \simfunccaller with its \param{Prototype}, \param{Input argument specifications}, and \param{Output argument specifications} parameters automatically populated. Note: \Simulink does not populate these fields automatically.
\item The syntactic interface (Section~\ref{sec:interface}) for a \Simulink module can be automatically generated. It can be represented visually in the model, or textually printed to the Command Window.
\item Module dependencies, such as \modelref, \library, and data dictionaries, can be detected by the tool, and summarized for the developer. This is useful for ensuring that the necessary definitions/files are available in order to compile and simulate. 
\item The four guidelines presented in Section~\ref{sec:guidelines} can be selected (\figurename~\ref{subfig:guideline}), automatically checked, and lists of violations are returned to the user.
\end{itemize}

The goal of this tool is to make it easier to create \Simulink modules in the first place, or to migrate existing models. An application of the tool is described in Section~\ref{sec:casestudy}.

\section{Nuclear Example}
\label{sec:casestudy}

\newcommand{\EP}{\key{Estimated\_Power}\@\xspace} 

This section describes how our concepts were applied to restructure a \Simulink implementation of a nuclear \gls{SDS}. The \gls{SDS} system senses whether conditions in the reactor are no longer safe, and controls the lowering of control rods to stop (``shutdown'') the reaction. With 605 subsystems, 74/7 top-level inputs/outputs and 6036 total blocks, the model is too large to be presented here. However, it represents the size of small to medium sized designs found in practice. 

Section~\ref{sec:example_module} describes how the proposed module structure was applied in the \gls{SDS} system. An explanation of how the \tool supports this process is provided in Section~\ref{sec:example_tool}. Section~\ref{sec:example_evaluation} goes on to evaluate the approach's effectiveness in terms of achieving a more modular system. The previously defined syntactic interface and guidelines are then applied to the example in Sections~\ref{sec:example_interface} and~\ref{sec:example_guidelines}.

\subsection{Application of the \Simulink Module Structure}
\label{sec:example_module}
\begin{figure}
\centering
	\begin{subfigure}[htb]{\columnwidth}
	\centering
		\includegraphics[width=\columnwidth]{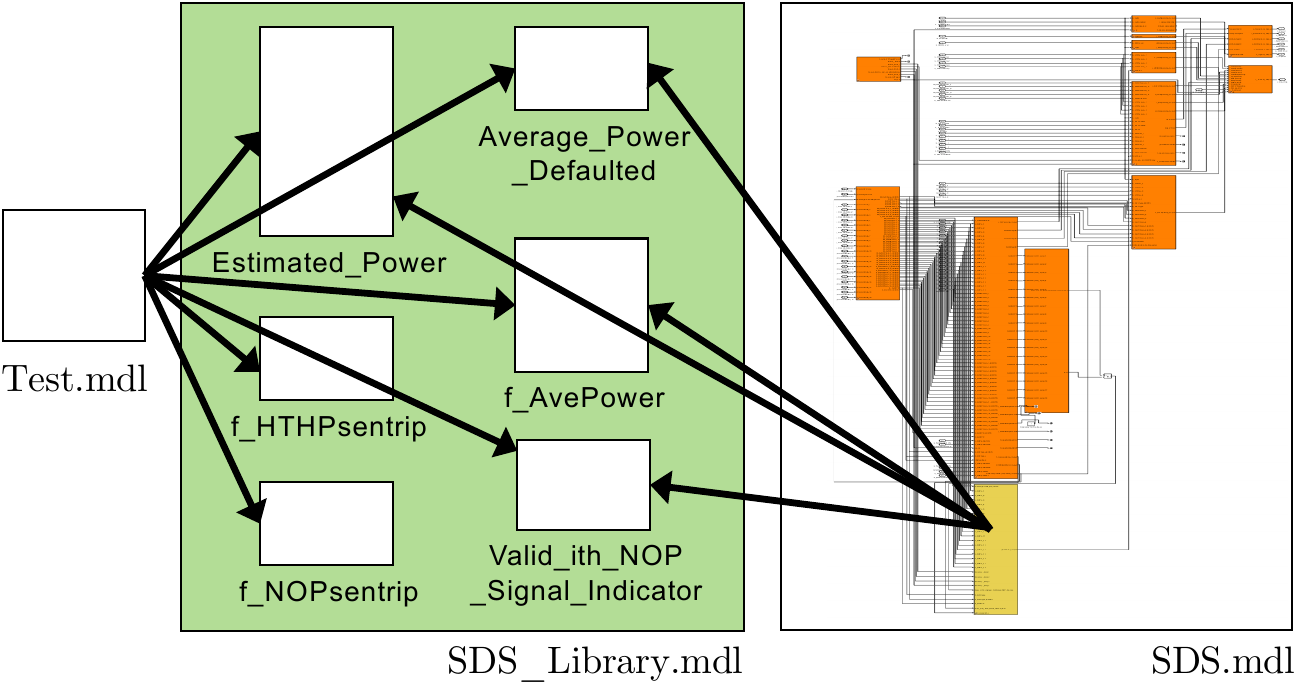}
		\caption{Before: \gls{PE} implemented in a \library.}
		\label{fig:example_before}
	\end{subfigure}
	\begin{subfigure}[htb]{\columnwidth}
	\centering
	\vspace{.25em}
  	\includegraphics[width=\columnwidth]{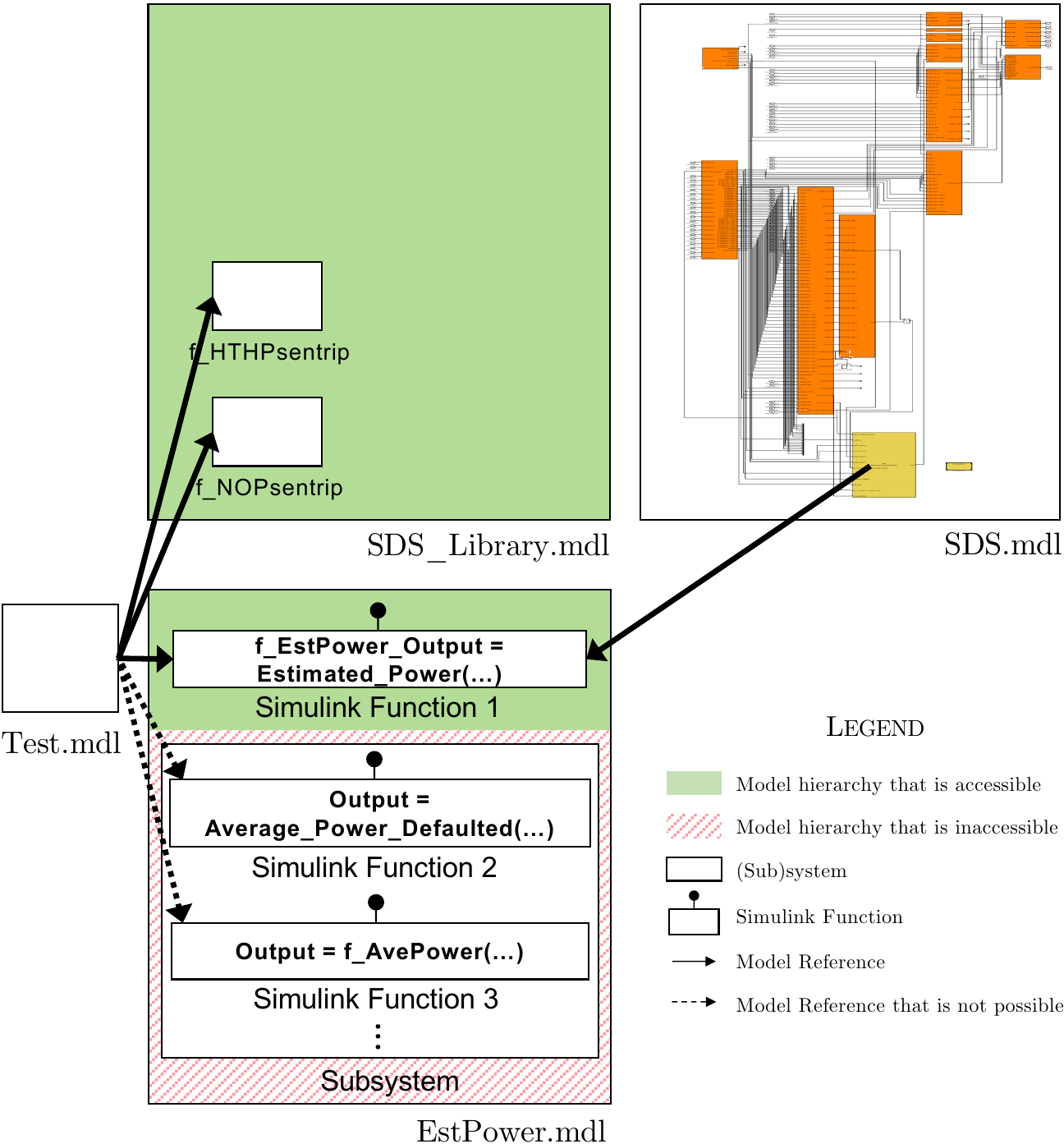}
  	\caption{After: \gls{PE} implemented with a module structure.}
  	\label{fig:example_after}
	\end{subfigure}
	\caption{Structure of the \gls{SDS} system.}
	\label{fig:example_system}
\end{figure}

The first \gls{SDS} implementation was implemented in \Matlab R2012a by other developers some years before \simfunc{s} were introduced in \Simulink. It heavily makes use of linked blocks in its design, which link to various blocks in the \gls{SDS} \library. The structure of the system is shown in \figurename~\ref{fig:example_before}. The \gls{PE} subsystem estimates the power of the reactor based on the average neutron over-power sensor values. It is one of the more complex components in the \gls{SDS}. The \gls{PE} implementation consists of several subsystems, which are defined in the \gls{SDS} \library. \gls{PE} contains secrets related to both hardware (\eg which sensors are used), and software (\eg how to accommodate for insufficient sensor readings). Unfortunately, a \library does not enforce information hiding~\cite{jaskolka2020comparison}. This is validated by the creation of a test model (\file{Test.mdl} in \figurename~\ref{fig:example_system}) to probe the \library. Any of the blocks in the \gls{SDS} \library can be used without restriction, and the internals of any subsystem are free to clients to use as well, even if this is not the developer's intent. It is not possible to selectively expose or hide functionality, as it is with our approach. 

By restructuring the \gls{PE} subsystems into a \Simulink module, we aim to hide implementation details  from users of the \gls{PE} module---\emph{the users should only be able to access the estimated power output value.} This is an essential difference between defining modules as we recommend compared to using ``coding'' guidelines that are not enforced by the language (\eg~\cite{jaskolka2020comparison}). \figurename~\ref{fig:example_after} shows the resulting module structure. A new model file (\file{EstPower.mdl}) was created and all related functionality was structured as a module as described in Section~\ref{sec:module}. This entailed organizing the operations as \simfunc blocks, choosing which are to be external and which are hidden in the module, and scoping them based on our guidelines (Section~\ref{sec:guidelines}). While there are many possible decompositions, the only exported function that is available for other modules to use is \EP. By placing it at the root level and setting the \param{\gls{FV}} parameter to \param{scoped}, it can be called like a member function (\ie \key{EstPower.\EP(\dots)}). The \gls{SDS} model imports this function definition using a \modelref to the module, and calls the function using a \simfunccaller block wherever the function is to be executed. Functionality in the \gls{SDS} \library unrelated to \gls{PE}, such as \key{f\_HTHPsentrip} and \key{f\_NOPsentrip}, remained as is.

\gls{SDV} was used to formally prove that the designs before and after restructuring were behaviourally equivalent. A verification harness was created that references both designs. We then instrumented the verification model with  proof objectives that state that each output of the original system must be equivalent to the corresponding output of the restructured system. \gls{SDV} is then executed to formally prove the specified properties. For the \gls{SDS} models, each property was successfully proven, thus verifying that the before/after systems are equivalent.

\subsection{Using the \tool}
\label{sec:example_tool}
The \tool facilitates the creation of \Simulink modules, as we have done in the \gls{SDS} example. The tool was used to make the changes described in Section~\ref{sec:casestudy}, generate syntactic interfaces, and check guidelines. In particular:

\begin{itemize}
\item The tool converted between the different kinds of scoping for \simfunc{s} when we were decomposing the \gls{PE} functionality into \simfunc{s}. The \simfunc{s} were easily made internal to the module, while the \key{f\_EstPower\_Output} was appropriately scoped so that it was an exported \simfunc and available to the \gls{SDS} model.
\item The tool assisted in calling the \key{f\_EstPower\_Output} \simfunc, with its appropriate qualifier, from the \gls{SDS} model. A \simfunccaller was automatically created with its \param{Prototype}, \param{Input argument specifications}, and \param{Output argument specifications} parameters pre-populated, saving time.
\item The syntactic interface for the \gls{PE} module was automatically generated. Its visual representation is shown and discussed in Section~\ref{sec:example_interface}. The interface information was used in the evaluation in Section~\ref{sec:example_evaluation} when examining changes to interface complexity.
\item Module dependencies were detected and listed by the tool. This information was used in the evaluation when analyzing interface complexity and coupling.
\item The four presented guidelines presented in Section~\ref{sec:guidelines} were automatically checked. This is further elaborated on in Section~\ref{sec:example_guidelines}.
\end{itemize}

In summary, the \tool saved much time in constructing the new \gls{PE} module, using the module's external function, and analyzing the module's interface, dependencies, and guideline compliance.

\subsection{Evaluation}
\label{sec:example_evaluation}
Through the use of our proposed \Simulink module structure, we aim to achieve designs that are robust with respect to change. In order to perform an evaluation of the proposed module structure, we sought to objectively quantify the improvement to modularity and information hiding. This was done by evaluating characteristics that are widely considered effective indicators of design structure and modularity of large systems, such as coupling and cohesion. In addition, we evaluated the approach by evaluating potential impacts to the designs in terms of complexity, structural coverage, and performance. 

\subsubsection{Information Hiding}
Although directly measuring information hiding has been attempted~\cite{rising1994information}, no metric has been widely accepted in either academia or industry. As a result, we use a qualitative analysis to reason about the effectiveness of our approach in supporting information hiding in the \gls{SDS} system. The \gls{PE} implementation contains secrets related to both hardware (\eg which sensors are used), and software (\eg how to accommodate for insufficient sensor readings). In the original design, knowledge of these secrets was easily leaked to the rest of the system, because the \gls{PE} implementation internals were accessible from the \gls{SDS} library without restriction. This was demonstrated through the use of a test model to probe the \gls{SDS} library (\figurename~\ref{fig:example_system}). Ultimately, the test model is able to access any of the elements in the library. For example, the \key{f\_AvePower} function can be used by the test model without restriction, even though this should be a hidden module as it will expose the secret of how the average is computed. Moreover, users can also directly create links to any blocks that are \emph{internal} to these top-level library blocks, further leaking internal design decisions. Restructuring the \gls{PE}-related subsystems into a \Simulink module allowed us to hide implementation details from users of the \gls{PE} module. In the newly modified design, information hiding is enforced so that the user can access \emph{only} the estimated power output value. With reference to Figs.~\ref{fig:example_before} and \ref{fig:example_after}, it is clear that the new module effectively hides the secrets that were previously accessible via \emph{Average\_Power\_Defaulted}, \emph{f\_AvePower}, and \emph{Valid\_ith\_NOP\_Signal\_Indicator}. In the restructured system, attempting to access any of these functions from the test model yields an error. This demonstrates that information hiding has improved.

\subsubsection{Interface Complexity}
\label{sec:interface_complexity}
\begin{figure}
\centering
	\begin{subfigure}[htb]{\columnwidth} 
	\centering
		\includegraphics[width=.7\columnwidth, trim={25em 35em 25em 15em}, clip]{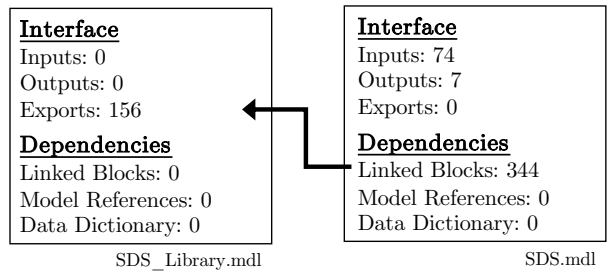}
		\caption{Before restructuring, the \gls{SDS} \library exposed all of its functionality, and the \gls{SDS} heavily depended on it.}
		\label{fig:example_interface_before}
	\end{subfigure}
	\begin{subfigure}[htb]{\columnwidth}
	\vspace{.25em}
	\centering
  	\includegraphics[width=.7\columnwidth, trim={25em 15em 25em 15em}, clip]{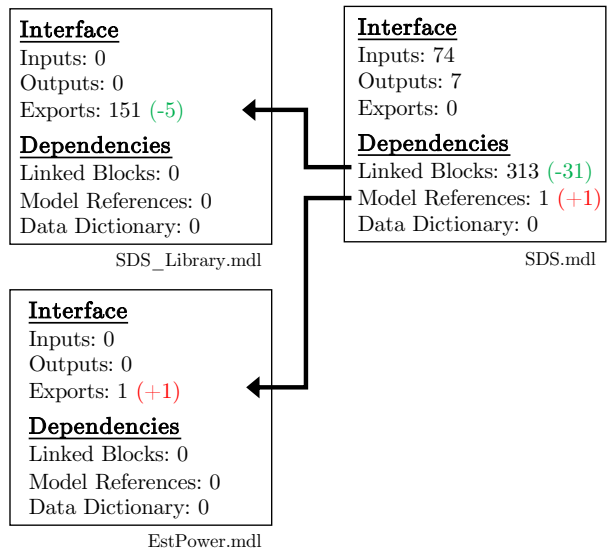}
  	\caption{After restructuring, the \gls{SDS} \library functionality related to \gls{PE} was hidden in EstPower, and only one function was exposed to the \gls{SDS}.}
  	\label{fig:example_interface_after}
	\end{subfigure}
	\caption{Interactions of the \gls{SDS} system.}
	\label{fig:example_interfaces}
\end{figure}

The complexity of a system is often attributed to the interactions, or interfaces, between the system's components. This complexity directly impacts the reusability, testability, and maintainability of the components. As discussed in Section~\ref{sec:interface}, the interface of a \Simulink model is generally considered to be comprised of \inport and \outport blocks, when in fact many other \Simulink elements also contribute to the interface. This is also reflected in the tools currently available for interface complexity checking, such as \Simulink Check Metrics Dashboard\footnote{\url{www.mathworks.com/help/slcheck/ug/collect-and-explore-metric-data-by-using-metrics-dashboard.html}} and \gls{MES} M-XRAY.\footnote{\url{www.model-engineers.com/en/quality-tools/mxray/}} As a result, we utilized the \Simulink module interface definition in Section~\ref{sec:interface} to provide a complete syntactic description of a module's interactions. The \tool automatically generates interface information, as well as dependency information, for a \Simulink module (Section~\ref{sec:example_tool}). Together this information provides an overall view of the many interactions that exist in a \Simulink system. The interactions for the nuclear example are shown in \figurename~\ref{fig:example_interfaces}. Each \Simulink model is represented with its interfaces and dependencies listed, while arrows represent the interactions. In \figurename~\ref{fig:example_interface_before}, the \gls{SDS} model was heavily coupled with the \gls{SDS} \library, with 344 links in its implementation to the 156 blocks exported from the \library. We restructured the \gls{PE} functionality into its own module, and five of the \gls{SDS} \library blocks were moved into this new module. We can see the \gls{SDS} \library in \figurename~\ref{fig:example_interface_before} is reduced from 156 blocks to 151 in \figurename~\ref{fig:example_interface_after}. Due to the \gls{PE} module's support for information hiding, only one of the blocks is actually exported by the module, and the remaining four are hidden and not available on the interface, causing a reduction of four fewer blocks on the interface. In total, these hidden blocks reduced the dependency of the \gls{SDS} model on the \library by 31 links. However, a new \modelref to include the interface of \file{EstPower.mdl} was introduced, as reflected in the \modelref count in the \gls{SDS} in \figurename~\ref{fig:example_interface_after}. Overall, the number of interactions in the system was reduced by 30, bringing down the total interface complexity.

\subsubsection{Coupling and Cohesion}
Coupling and cohesion are well known as indicators of the quality of a program decomposition, and are related to the concept of information hiding~\cite{stevens1999structured}. Coupling is a measure of the interconnections between modules, and increases as the complexity of the interfaces between modules increase~\cite{stevens1999structured}. While information hiding aims to hide the implementation details of a module, coupling measures how much another module is reliant on another module. Minimizing the coupling of a module makes it more robust with respect to changes because it reduces the connections by which changes and errors can propagate~\cite{stevens1999structured}. Thus, a good design that implements information hiding will also exhibit low coupling. Designs with low coupling and high cohesion lead to software that is more reliable and more maintainable~\cite{fenton2014software}. Cohesion is a measure of the relationships of the elements \emph{within} a module, with the aim of ensuring that module elements are highly related to each other~\cite{stevens1999structured}. Cohesion also supports information hiding by ensuring that the contents of a module are strongly related to one secret. In the context of \Simulink, coupling and cohesion are typically defined on a single \Simulink model, based on the interactions of the contained blocks~\cite{plaska2011simulink}, or specifically \subsystem blocks~\cite{gerlitz2016architectural,dajsuren2013simulink}. There is a lack of system-level metrics for coupling and cohesion in the \Simulink environment, and in turn, an absence of tools that automatically measure these qualities. Nevertheless, we manually analyzed the impact to coupling and cohesion that our approach had on the \gls{SDS}. 

The interconnections between the \gls{SDS} \library and the \gls{SDS} model as they were originally structured is shown in \figurename~\ref{fig:example_before} via the arrows that exist between the two. Although only a few of all of the connections are drawn in the figure, it is evident that the \gls{PE} implementation in \file{SDS.mdl} was highly coupled with \file{SDS\_Library.mdl} when compared with the restructured design in \figurename~\ref{fig:example_after}. The second design reduced the interconnections to only one interconnection between \file{SDS.mdl} and \file{EstPower.mdl}. This reduction in coupling was additionally observed when comparing the syntactic interfaces of the models, as described in the previous section and shown in \figurename~\ref{fig:example_interfaces}.

In the original system, the \gls{SDS} library contained all functionality related to the \gls{SDS}. In the restructured system, those functions related to \gls{PE} were decomposed into a separate module. This resulted in a more cohesive module of more closely related functions, in comparison to the original. Therefore, our approach positively impacted the design by reducing coupling and making the module more cohesive.

\subsubsection{Cyclomatic Complexity}
\label{sec:cyclomatic_complexity}
Cyclomatic complexity is the most widely-used metric in industry for gauging the structural complexity of software. Cyclomatic complexity measures the amount of decision logic in a program, or more specifically, the number of linearly independent execution paths through a program~\cite{mccabe1976complexity}. It has been shown to be indicative of the maintainability of software, which is our ultimate goal in supporting information hiding and modularity. MathWorks has adapted this metric for use directly on \Simulink models, and is provided as an architecture metric via the \Simulink Check toolbox. For more information on how MathWorks adapts this metric to \Simulink models, please see the \Simulink Check Reference~\cite{mathworks2019simulink}. 

We leveraged \Simulink Check to automatically compute cyclomatic complexity values for the nuclear \gls{SDS} \Simulink model. \tablename~\ref{tbl:commplexity} shows the cyclomatic complexity results before and after the restructuring of the \gls{PE} functionality into a \Simulink module. The \gls{PE} functionality originally had a cyclomatic complexity of 143. The \gls{SDS} model decreased in cyclomatic complexity by 143. We expected this reduction in complexity as a result of moving the \gls{PE} functionality out of \gls{SDS} into its own module, but wanted to check that such a change did not inadvertently increase the complexity in \gls{SDS}. Due to the application of our proposed module structure, the \gls{PE} module itself had a substantial reduction in cyclomatic complexity by approximately 43\%.

\begin{table}[!htbp]
	\centering
	\caption{Cyclomatic complexity comparison.}
	\label{tbl:commplexity}
	\begin{tabular}{lccc}
		\hline
					   & Before	& After	& Difference  \\ \hline \hline
		SDS 		 & 2413 	& 2270	& -143 (6\%)	\\ \hline
		EstPower & 143	  & 81		&	-62	(43.4\%)	\\ \hline
	\end{tabular}
\end{table}

The change in \gls{PE} cyclomatic complexity is a result of our module structure leveraging \simfunc blocks, whereas the original design relied heavily on \library blocks containing virtual \subsystem{s}. In the \gls{PE} functionality shown in \figurename~\ref{fig:example_before}, the \key{Valid\_ith\_NOP\_Signal\_Indicator} \library block was used/linked 18 times. When the \gls{PE} functionality was separated into a separate \Simulink module, this block was converted to a \simfunc block in order to hide its implementation according to our proposed \Simulink module structure. In previous work, we performed a thorough comparison between componentization constructs, including \library and \simfunc blocks~\cite{jaskolka2020comparison}. This analysis showed that when virtual (\ie non-atomic) \subsystem{s} are placed in a Library, each instance of the linked block is separate. Ultimately, a \subsystem in a \library will result in non-reusable code, that is, it will not be generated into a function. The original design contained 18 links to the \key{Valid\_ith\_NOP\_Signal\_Indicator} block, and each one added a cyclomatic complexity value of 4 per block (for a total of 72). Converting the linked \library block to a \simfunc and then calling it 18 times reduced the cyclomatic complexity to a value of 4, as the \simfunc provides one reusable definition of \key{Valid\_ith\_NOP\_Signal\_Indicator}, for a total cyclomatic complexity reduction of 68. However, the use of \simfunc{s} can also increase the cyclomatic complexity when converting a virtual \subsystem to a \simfunc. MathWorks has adapted the cyclomatic complexity metric to the \Simulink language such that each atomic \subsystem adds a value of 1 to the complexity~\cite{mathworks2019simulinkc}. During the restructuring, 5 \simfunc{s} replaced \subsystem blocks, and as a result, the cyclomatic complexity increased by 5. Moreover, the base cyclomatic complexity for a model with decision points is 1, so the new model also adds a value of 1 to the cyclomatic complexity of the \gls{PE} module. Ultimately, the new structure and use of \simfunc{s} resulted in the cyclomatic complexity being reduced by 68, and then increased by 6, for a net reduction of 62, as shown in \tablename~\ref{tbl:commplexity}. 

In summary, although the use of \simfunc blocks may come at a small cost to cyclomatic complexity, they produce more reusable designs while also supporting information hiding through their scoping ability. It is also important to note that \simfunc blocks are a type of atomic subsystem, meaning that their contents are executed as a single unit. When converting from a virtual (or non-atomic) \subsystem, the execution order may be impacted. To remedy this, \Simulink provides the ability to assign priorities to nonvirtual blocks to change their execution order according to the needs of the design~\cite[ch.37]{mathworks2019simulink}.

\subsubsection{Testability}
The \gls{SDV} toolbox automatically generates test cases for \Simulink models in order to maximize the structural coverage metrics of decision, condition, \gls{MCDC}, and execution coverage. \gls{SDV} was run on each system for approximately 20 hours each, and the results are reported in \tablename{s}~\ref{tbl:objective} and~\ref{tbl:coverage}. There were 80 fewer total test objectives in the restructured system, as shown in \tablename~\ref{tbl:objective}. Overall, the number of processed objectives slightly increased. The number of satisfied objectives stayed relatively the same, while slightly more objectives were decided to be unsatisfiable by the prover. Also, the number of undecided objectives slightly decreased. 

\tablename~\ref{tbl:coverage} reports no significant changes to structural coverage in the totals for the whole system. In measuring the structural coverage for the \gls{PE} module alone, there was no change in the condition, \gls{MCDC}, or execution coverage. However, a reduction of 10\% in decision coverage was observed, for the same reasons outlined in Section~\ref{sec:cyclomatic_complexity}. This change was primarily due to the \key{Valid\_ith\_NOP\_Signal\_Indicator} \library block being converted to a \simfunc, making it one reusable function definition instead of 18 separate definitions. This eliminated 85 decision objectives, 68 of which were satisfied. The addition of the 5 \simfunc blocks themselves increased the decision objectives by 5, all of which were satisfied, thus slightly increasing decision coverage. This was to be expected as the MathWork's definition of decision coverage encompasses function call execution. However, the net decision coverage and decision objectives were decreased by 63 and 80, respectively, simply due to the elimination of function clones. 

\begin{table}[!htbp]
	\centering
	\caption{Test objective comparison for the whole \gls{SDS} system.}
	\label{tbl:objective}
	\begin{tabular}{lllc}
		\hline
					  						& Before 				 & After 					 	& Relative Difference   \\ \hline \hline
		Total Objectives 		& 7099	 				 & 7019	  			 		& $-$1.1\%			\\ 
		Processed 					&	4427 (62.4\%)  & 4705	(67\%)	 		& +4.6\%			\\ 
		Satisfied						& 2899 (40.8\%)  & 2836	(40.4\%)	 	& $-$0.4\%			\\ 
		Unsatisfiable				& 1528 (21.5\%)  & 1869 (26.6\%)		& +5.1\%			\\ 
		Undecided						& 2672 (37.6\%)  & 2314 (33\%) 			& $-$4.6\%			\\ \hline
	\end{tabular}
\end{table}

\begin{table}[!htbp]
	\centering
	\caption{Structural coverage comparison for the whole \gls{SDS} system and \gls{PE}.}
	\label{tbl:coverage}
	\begin{tabular}{lccc|ccc}
		\hline
								  & \multicolumn{3}{c|}{Total} & \multicolumn{3}{c}{EstPower} \\ \hline 
					  			& Before 	& After & Difference  & Before &  After & Difference \\ \hline \hline
		Decision  		& 61\%		& 60\% 	& $-$1\%	& 68\% (111/163)  & 58\% (48/83) 		& $-$10\%	\\ 
		Condition			&	36\%  	& 36\% 	& 0\%			& 30\%	& 30\% 		&	0\%		\\ 
		MCDC		 			&	15\%  	& 15\% 	& 0\%			& 4\%		& 4\% 		&	0\%		\\ 
		Execution			& 99\% 		& 99\% 	& 0\%			& 100\% & 100\%		&	0\%		\\ \hline
	\end{tabular}
\end{table}

\subsubsection{Performance Comparison}
As our approach relies heavily on the use of \simfunc blocks, it is important to be cognizant of the potential for added overhead due to the increase in function calls and switching between modules in the new decomposition~\cite{parnas1972criteria}. To determine whether there was a change in efficiency between the original system and our modified system, they were simulated 200 times each. Our decomposition exhibited a minor increase in the  execution times, as shown in \tablename~\ref{tbl:executiontime}. On average there was a penalty of 1.4\%, but it is to be expected due to the introduction of function calls. 

\begin{table}[!htbp]
	\centering
	\caption{Model execution time comparison.}
	\label{tbl:executiontime}
	\begin{tabular}{lcccc}
		\hline
					  & Before (\textit{ms}) 	& After (\textit{ms}) & Difference & Percent Difference \\ \hline \hline
		Mean 		& 5770	& 5848 	& +78 	& +1.4\%		\\ 
		Best 		&	5719  & 5798	& +79		& +1.4\%		\\ 
		Worst		& 7811  & 8116	& +305 	& +4.0\%		\\ \hline
	\end{tabular}
\end{table}

\subsubsection{Summary of the Evaluation}
Analyzing the modularity of a system entails measuring qualities such as cohesion, coupling, complexity, and others. In our evaluation, we looked at some of the best known indicators of modularity. We observed that the restructured system had an increase in information hiding, due to the appropriate scoping of design secrets. There was also a decrease in interface complexity, as the new decomposition hid previously exposed internal functions, thus removing them from the interface. This in turn resulted in a decrease in coupling, as the system interactions were reduced. An increase in cohesion was observed due to the grouping of \gls{PE}-specific functionality. The cyclomatic complexity of the \gls{SDS} decreased due to the decomposition of the system, and the \gls{PE} module experienced a decrease in cyclomatic complexity due to the use of \simfunc{s} for reusable functionality. In terms of structural coverage, although there was a minor reduction in the number of coverage objectives, there were no significant changes to coverage metrics. However, a slightly lower decision coverage in the \gls{PE} module was exhibited due to \simfunc{s} eliminating clones. As expected, there was a small performance penalty of 1.4\% on average, due to \simfunc overhead. In summary, the use of the proposed module concept for decomposing a \Simulink model objectively exhibits improvements in key qualities that are indicative of modularity. The application of our proposed approach lead to more a modular design overall.

\subsection{Using the Module Interface}
\label{sec:example_interface}
\begin{figure} 
\centering
	\begin{subfigure}[b]{\textwidth}
	\centering
	\includegraphics[width=\textwidth]{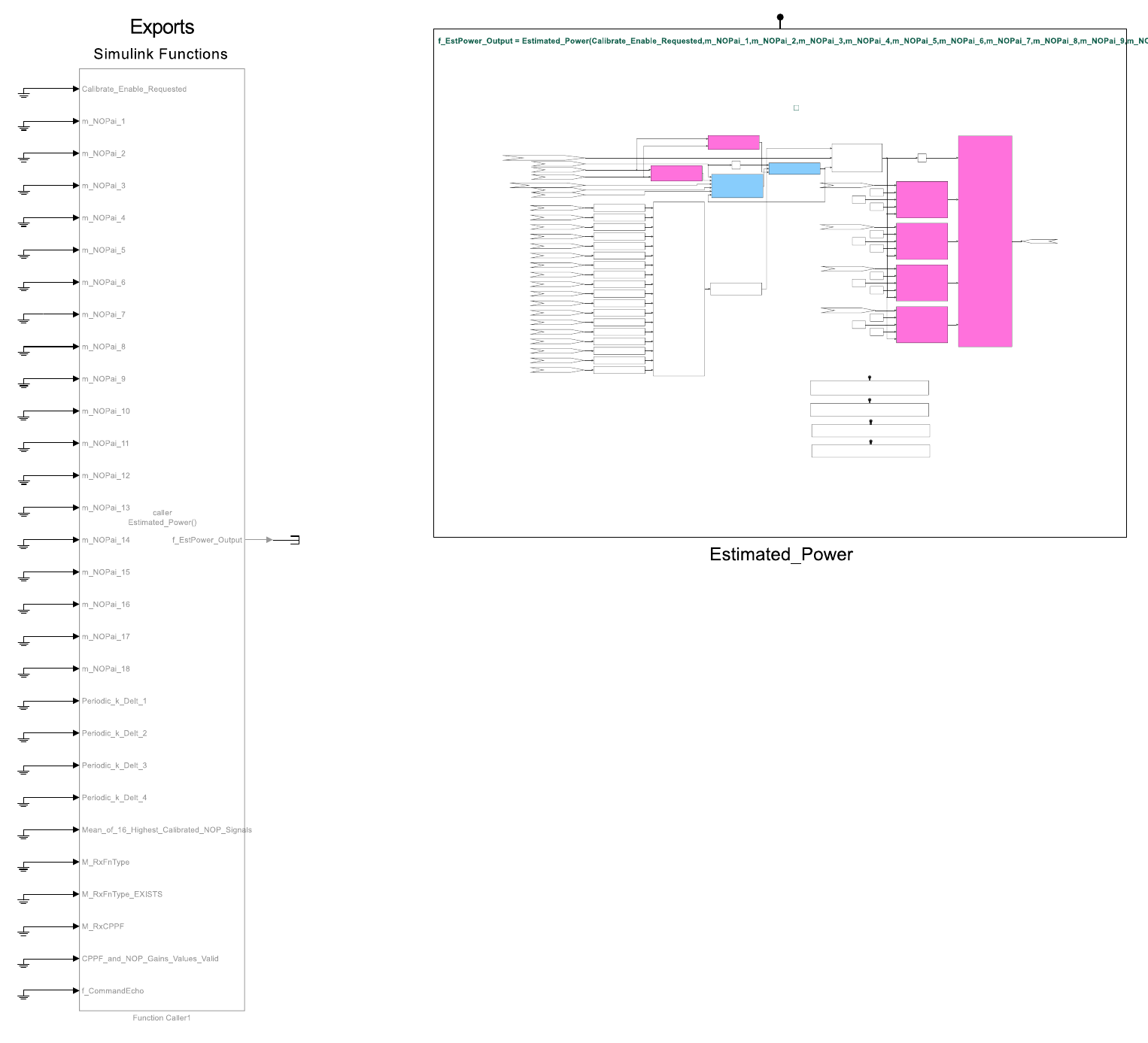}
	\caption{The \tool interface showing one exported \simfunc.}
	\label{fig:pe_moduletool}
	\end{subfigure}
	
\vspace{1em}

	\begin{subfigure}[b]{\textwidth}
	\centering
	\includegraphics[width=.65\textwidth]{figs/NuclearExample/InterfaceDisplay_EstPower}
	\caption{The MathWorks Interface Display does not show any exported \simfunc{s}.}
	\label{fig:pe_interfacedisplay}
	\end{subfigure}
  	\caption{\gls{PE} module interface representations.}
  	\label{fig:pe_interface}
\end{figure}

In Section~\ref{sec:interface_complexity} the syntactic interfaces of the \gls{SDS} and \gls{PE} models were generated with the \tool. \figurename~\ref{fig:pe_interface} shows the generated interface representations from both the \tool and the MathWorks Interface Display for the \gls{PE} module. The representation of the interface should show that a single \simfunc is exported from \gls{PE}, as also depicted in Figs.~\ref{fig:example_after} and~\ref{fig:example_interface_after}. 
In the interface generated by the \tool shown in \figurename~\ref{fig:pe_moduletool}, we can see the  \key{Estimated\_Power} function under the ``Exports'' heading, to the left of the implementation. Unfortunately, the exported function is not shown in the MathWorks Interface Display view shown in \figurename~\ref{fig:pe_interfacedisplay}, because it only shows the \inport and \outport{s}, of which there are none in this module. Our definition of an interface promotes a better, more complete, view of elements present on the interface, as demonstrated by the visible exported \simfunc present in \figurename~\ref{fig:pe_moduletool}, but missing from \figurename~\ref{fig:pe_interfacedisplay}.
Although this is a simple example, it indicates that our definition of a \Simulink module interface ultimately leads to a better understanding of the data flow into and out of the model. 

\subsection{Using the Guidelines}
\label{sec:example_guidelines}
The new design adheres to the guidelines presented in Section~\ref{sec:guidelines}, as follows. All functions were placed as low as possible in the module such that any corresponding \simfunccaller blocks in the \gls{SDS} module can still access them (Guideline 1). This minimizes the scope of each of the \simfunc{s}, and restricts their accessibility in the module to only where it is required. Only the \EP function is exported in order to make this functionality available on the \gls{PE} interface, so that the \gls{SDS} model can import it. All \simfunc \param{\gls{FV}} parameters are set to \param{scoped}, as global visibility is not needed in the system (Guideline 2). These guidelines helped to enforce information hiding in the system, and ensured the interfaces were as minimal as possible. Furthermore, each \simfunc has a unique name, resulting in no shadowing (Guideline 3). This made the new decomposition of the \gls{PE} module easy to understand. Lastly, the \gls{PE} module's functionality was already prepared as a production model, and did not contain any constructs that utilized the base workspace (Guideline 4). No further action was needed to support this guideline. The \tool was used to automatically validate that the \gls{PE} module complied with these guidelines. 

\section{Conclusions and Future Work}
\label{sec:conclusions}
Due to the increasing complexity of models, and their position as primary design artifacts maintained over many years, it is important to support modularity principles in \Simulink. In other languages, modularity has been supported by following principles for ensuring modular decomposition, hiding implementation details, and separating the interface from implementation. To understand how to adapt these practices for \gls{MBD} with \Simulink, this paper presented a comparison of \Simulink constructs to the constructs of the C language. Then, a novel approach for modular modelling was proposed and described. The approach entails structuring models as modules to support encapsulation and facilitate information hiding. The definition of a module interface was given, which effectively represents all data flow across the module boundary. Four new guidelines to encourage best practices were presented. The \tool was developed to automate the aforementioned contributions, and made available as an open-source contribution to the \Matlab Central File Exchange. This paper also presented a proof of concept application of these ideas on an industrial system from the nuclear domain, which demonstrated the feasibility of the overall approach. We evaluated the approach and demonstrated that the restructured system increased in support for information hiding, simplified interfaces, decreased module coupling and increased cohesion, while not increasing the cyclomatic complexity of the system. Although decision coverage decreased, this was attributed to \simfunc blocks replacing \library blocks that caused clones, which inflated the decision coverage objectives in the first place. Condition, \gls{MCDC}, and execution coverage remained largely the same. A small performance penalty of 1.4\% was incurred due to the use of \simfunc blocks, but this is a small price to pay for the added modularity and ability to enforce information hiding in the system. In summary, this paper provided a new approach to structuring modular \Simulink software systems.

As future work, we plan to extend this approach to the Stateflow environment, which is a subset of the \Simulink language that uses functions in state machines and flow charts. A survey of likely changes in industrial models is underway. Different versions of industrial software from a version control system are being analyzed to create a taxonomy of likely changes for \Simulink models. This will help in understanding how frequently occurring changes can be better accommodated in \Simulink, as well as potentially identify other software engineering principles which can prove useful. A comprehensive review of other traditional software engineering principles that need to be better supported in \Matlab \Simulink is necessary, followed by further work towards addressing any identified gaps. 

\bibliographystyle{spmpsci}
\bibliography{modularity}

\end{document}